\documentclass[sigplan,10pt]{acmart}

\renewcommand\footnotetextcopyrightpermission[1]{} 
\usepackage{xcolor}
\newcommand{\cmark}{\textcolor{green!55!black}{\ding{51}}}
\newcommand{\xmark}{\textcolor{red!70!black}{\ding{55}}}

\usepackage{booktabs,multirow,tabularx,makecell,array}
\usepackage{xspace}        
\usepackage{ragged2e}
\usepackage{graphicx}   

\usepackage{amsmath,amssymb}

\usepackage{tikz}

\usepackage{algorithm}      
\usepackage{algpseudocode}
\algtext*{EndFor}
\algtext*{EndWhile}
\algtext*{EndIf}
\algtext*{EndProcedure}
\algtext*{EndFunction}

\usepackage{mathtools}
\usepackage{pifont}
\usepackage{pgfplots}
\pgfplotsset{compat=1.18}

\newcolumntype{Y}{>{\RaggedRight\arraybackslash}X}

\usepackage{siunitx}
\usepackage{amsthm}          

\newcommand{\sys}{SlackServe\xspace}

\newcommand{\marklessfootnote}[1]{
  \begingroup
  \renewcommand\thefootnote{}
  \footnote{#1}
  \addtocounter{footnote}{-1}
  \endgroup
}

\usepackage{soul}

\begin{document}

\title{Adaptive Resource Management and Quality Control for Streaming Video Generation}

\author{
  \fontsize{11.5}{12.5}\selectfont{Yifei Xia, Hao Yuan, Suhan Ling, Haoran Sun, Hanke Zhang, Xupeng Miao, Fangcheng Fu, Bin Cui}\\
  {\fontsize{9.5}{12.5}\selectfont\emph{The Hetu Team @ Peking University}}
}
\pagestyle{plain}

\begin{abstract}
Autoregressive diffusion transformers (AR-DiTs) recast video generation from an offline paradigm to a real-time streaming one: the model generates video one chunk at a time, making each chunk available for playout once produced. The \textit{service-level objective} (SLO) for this paradigm is no longer fixed latency or throughput, but \textit{the preservation of playout continuity}: generation must stay ahead of the playout timeline. Once generation falls behind, the remaining playable buffer (\textit{playout slack}) is exhausted, and users experience visible stalls. This objective reveals two serving design insights. First, real-time video generation has a dynamic SLO that evolves with playout progress, so resources should move toward streams with lower playout slack. Second, an acceptable chunk delivered on time is preferable to a late high-fidelity chunk, so per-chunk fidelity configurations should adapt to available playout slack.
Guided by these insights, we present \sys, a playout-slack-driven serving system that preserves playout continuity in real-time streaming video generation. \sys uses playout slack as a unified signal, reallocating resources across streams through three-tier priority queues, re-homing, and elastic sequence parallelism, while selecting per-chunk fidelity configurations within each stream through \textit{Bi-Modal Pareto Routing} under a quality floor. On a 16-H100 GPU cluster, \sys improves \textit{Quality of Experience} (QoE), measured by \textit{Continuous Play Ratio} (CPR), by 1.64$\times$--3.29$\times$ and reduces \textit{Time to First Chunk} (TTFC) by 1.61$\times$--9.65$\times$ over baselines, while preserving comparable generation quality.
\end{abstract}

\maketitle


\marklessfootnote{Contact: Yifei Xia (yifeixia@stu.pku.edu.cn), Fangcheng Fu (ccchengff@sjtu.edu.cn) and Bin Cui
(bin.cui@pku.edu.cn)} 

\section{Introduction}

\begin{figure}[t]
\centering
\includegraphics[width=\linewidth]{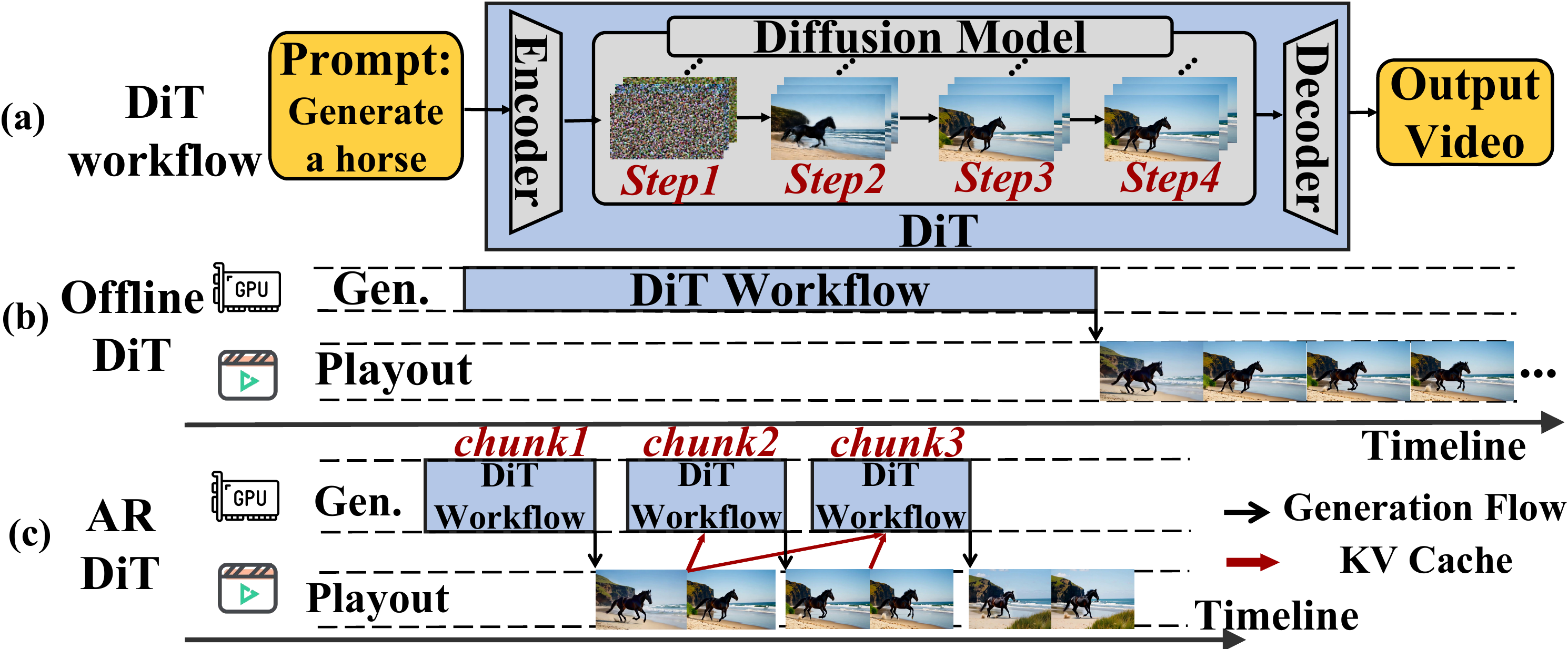}
\caption{Workflow and generation paradigms of DiT-based video generation.}
\label{fig:p1}
\end{figure}

Recent advances in video generation are pushing the field beyond offline, one-shot generation~\cite{hunyuanvideo,cogvideox,wan} toward autoregressive and interactive generation paradigms~\cite{selfforcing,causalforcing,magi,helios}. Offline diffusion transformers (DiTs)~\cite{ddim,ddpm,peebles2023scalable}, as shown in Figure~\ref{fig:p1}(a) and (b), generate all frames through one multi-step denoising process over the entire video. This paradigm incurs high latency because denoising long video sequences is time-consuming, often taking tens of minutes to generate a 5-second, 81-frame video~\cite{adaspa}. 
Thus, it can serve latency-insensitive scenarios where users can wait~\cite{chen2024videocrafter2,wang2025lavie}, but falls short for interactive or real-time applications~\cite{causvid,streamdit}.

Autoregressive diffusion transformers (AR-DiTs)~\cite{selfforcing,causalforcing,selfforcing++,rollingforcing,magi,helios}, represented by Self-Forcing~\cite{selfforcing} and Causal-Forcing~\cite{causalforcing}, recast video generation as a chunk-wise streaming process. As shown in Figure~\ref{fig:p1}(c), the model generates video one chunk at a time, making each chunk available for playout once produced. For each chunk generation, the model uses the KV cache of previous chunks to preserve temporal coherence~\cite{selfforcing,ca2vdm}. 
By shortening the input sequence length, chunk-wise generation reduces generation time and, more importantly, makes real-time streaming possible~\cite{streamdit,streamingt2v}. This capability enables emerging applications, such as AI live streaming and real-time visual effects~\cite{motionstream,streamdiffusion,streamdiffusionv2}.

\begin{figure}[t]
\centering
\includegraphics[width=\linewidth]{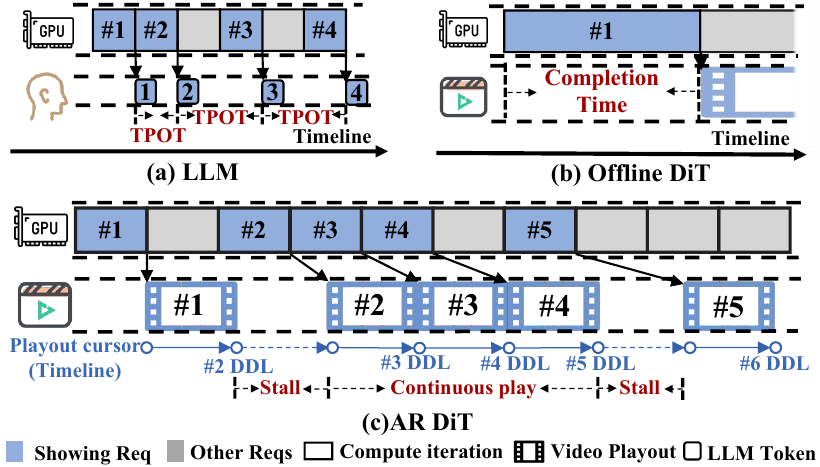}
\caption{Representative SLOs across LLM, offline DiT, and AR-DiT serving. \#i denotes the $i$-th compute iteration, which produces one token in LLMs and one chunk in AR-DiTs.}
\label{fig:p2}
\end{figure}

Despite their promise, efficient serving for real-time streaming video generation remains a pressing yet underexplored system problem~\cite{streamdiffusion,streamdiffusionv2,streamdit}.
The core challenge is an objective mismatch: existing serving abstractions     often center on fixed latency or throughput targets~\cite{vllm,orca,distserve,diffserve,tridentserve,sarathi-serve}, which are insufficient for this streaming setting. As shown in Figure~\ref{fig:p2}, the primary \emph{service-level objective} (SLO)~\cite{srebook} for each video stream is no longer a fixed latency target, such as \emph{Time per Output Token} (TPOT)~\cite{distserve} in large language models~\cite{gpt3,transformer} (LLMs) or completion time in offline DiTs~\cite{tridentserve}, but \textit{preserving user-visible \textbf{playout continuity}}. To be specific, the generated video must stay ahead of the playout timeline.
If generation falls behind, the remaining playable buffer is exhausted, causing users to experience playout stalls.
We define this remaining playable buffer as \textit{\textbf{playout slack}}.

This objective exposes two key serving properties that existing serving systems do not account for.

\begin{figure}[t]
\centering
\includegraphics[width=\linewidth]{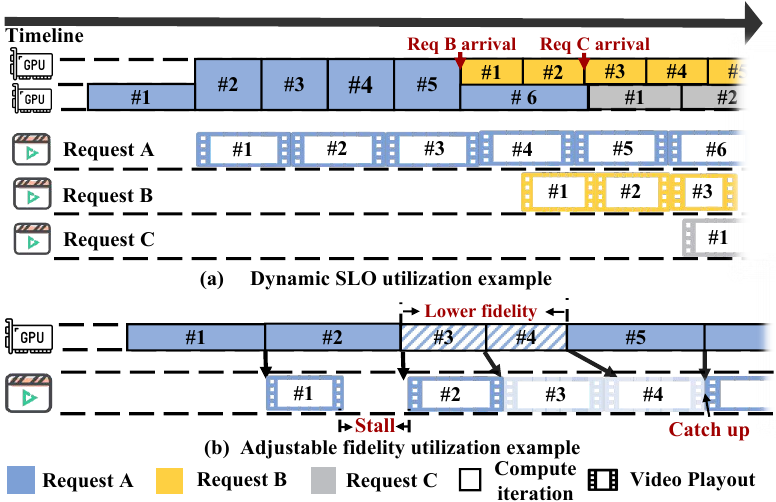}
\caption{Examples that utilize the properties of AR-DiT.}
\label{fig:p3}
\end{figure}

\underline{\textit{Property 1:}} \textit{Real-time video generation has a \textbf{dynamic SLO} that evolves with the playout timeline.} As shown in Figure~\ref{fig:p2}(c), unlike fixed TPOT-style targets in LLM serving, a real-time video stream\footnote{A video stream is a real-time video generation session.}'s SLO threshold changes after admission. Its SLO evolves as playout advances: the playout cursor moves at $fps$ frames per second, continuously updating the effective deadline (DDL) of the next chunk. This makes resource allocation decisions and SLO evolution mutually dependent. In other words, resource allocation changes a stream's playout slack, while playout slack in turn guides how compute should be allocated\footnote{Playout slack can also change due to user behavior, including prompt switches that reset it and playout pauses that let it accumulate.}.
For example, as shown in Figure~\ref{fig:p3}(a), when the system load is light (only stream A is active), it can allocate additional compute resources, such as increasing parallelism, to accumulate playout slack. When load increases (streams B and C arrive) and compute becomes scarce, stream A remains non-urgent due to accumulated slack, allowing resources to be allocated to streams B and C as they are under heavier playout pressure.
Such dynamic SLO control cannot be captured by a fixed metric like TPOT.

\underline{\textit{Property 2:}} \textit{Real-time video generation has \textbf{adjustable fidelity} due to a deadline-sensitive quality-latency tradeoff: timely chunks can outweigh late high-fidelity chunks.} 
\textit{Playout continuity} can be tracked by an objective runtime quantity: playout slack.
When playout slack is exhausted, \textit{playout continuity} is violated and users immediately perceive stalls.
Thus, when system load is heavy, it is acceptable to temporarily deliver a chunk with lower fidelity on time, rather than a high-fidelity chunk late.
The same principle is widely used in deadline-sensitive scenarios, such as adaptive-bitrate streaming over pre-encoded segments (e.g., Pensieve, BBA, and MPC)~\cite{pensieve,bba,mpc}, and dynamic resolution scaling for game rendering~\cite{furion}.
Fortunately, the chunk-wise generation pattern of AR-DiTs allows the serving system to select a fidelity configuration for each chunk based on current playout slack. As exemplified in Figure~\ref{fig:p3}(b), the system can temporarily choose lower-cost fidelity configurations when playout slack is running out, rather than fixing stream-wide quality.

These properties open a design space largely unexplored by existing serving paradigms: resources must be reallocated across streams, and fidelity must be adjusted across chunks according to the real-time distribution of playout slack (\S\ref{subsec:motivation}).

While these properties open a new design space, realizing them in a stable serving system raises two challenges.

\underline{\textit{Challenge 1.}} \textit{Playout slack dynamics make resource allocation decisions difficult.} 
At the algorithm level, resource allocation and slack evolution continuously affect each other (\S\ref{sec:resource}). 
Therefore, making ad-hoc resource allocation decisions at runtime is challenging in this interdependent decision space.
At the system level, AR-DiTs are stateful, typically through KV cache~\cite{magi,selfforcing}, so reallocating compute often requires state migration and introduces overhead.

\underline{\textit{Challenge 2.}} \textit{Multidimensional fidelity knobs complicate the speed--quality tradeoff.} Unlike adaptive-bitrate streaming~\cite{bba}, which mainly changes transmission bitrate, AR-DiTs expose multiple speed--quality tradeoff knobs, such as sparse attention and quantization~\cite{karras2022elucidating,sageattention,dmd,lightforcing,awq}, whose combinations create a complex decision space (\S\ref{sec:backvideogen},~\S\ref{sec:precision}). 
As a result, fidelity-configuration selection depends on complex interactions among workload state, serving pressure, and quality requirements, making the tradeoff non-trivial (\S\ref{sec:precision}). 

To address these challenges, we present \textit{\textbf{\sys}}, a playout-slack-driven serving system for real-time streaming video generation. To our knowledge, \sys is the first system built around \textit{dynamic SLOs} and \textit{adjustable fidelity} to preserve \textit{playout continuity} in real-time video generation.

To handle dynamic SLOs, \sys introduces \textit{Slack-Driven Resource Reallocation} (\S\ref{sec:resource}), which combines two common resource-reallocation mechanisms: preemption and compute expansion. Specifically, \sys uses three-tier priority queues (\S\ref{sec:queue}) and cross-worker re-homing (\S\ref{sec:rehome}) for local and remote preemption, respectively, and uses elastic sequence parallelism~\cite{ulysses,ring,loongserve} (elastic SP) (\S\ref{sec:elastic}) for compute expansion on urgent streams.
\sys jointly coordinates these mechanisms with adjustable fidelity, composing and triggering them in appropriate orders to make reallocation decisions in the interdependent decision space.

To enable adjustable fidelity, \sys introduces \textit{Slack-Driven Fidelity Selection} (\S\ref{sec:precision}). It first establishes a profiled Pareto frontier~\cite{halpern2019one,miettinen1999nonlinear} over the fidelity-configuration space, capturing the non-dominated speed--quality tradeoffs among candidate fidelity configurations. Based on this frontier, \sys designs \textit{Bi-Modal Pareto Routing (BMPR)}, which routes chunks to suitable fidelity configurations according to the current playout-slack budget while enforcing a global quality floor through a bi-modal decision rule.

At the system level, \sys realizes these mechanisms through three decoupled asynchronous planes (\S\ref{sec:three-planes}). The \textit{Control Plane} jointly makes resource allocation and fidelity selection decisions based on playout slack. The \textit{Execution Plane} dispatches streams and applies the \textit{Control Plane}'s decisions. The \textit{State Plane} manages the paged KV cache and performs asynchronous migration according to these decisions. This design decouples decision making, execution, and state migration while overlapping migration with ongoing chunk computation whenever possible.

We summarize our main contributions as follows.
\begin{itemize}
    \item We present \sys, to our knowledge the first serving system designed to preserve playout continuity in real-time video generation by exploiting two properties of streaming generation: dynamic SLOs and adjustable fidelity.
    \item Building on dynamic SLOs, we design \textit{Slack-Driven Resource Reallocation}, which combines three-tier priority queues, re-homing, and elastic SP to mitigate resource mismatch and direct compute toward streams at risk of stalling.
    \item Building on adjustable fidelity, we design \textit{Slack-Driven Fidelity Selection}, which uses Bi-Modal Pareto Routing (BMPR) to select a Pareto-optimal fidelity configuration for each chunk under a time budget and quality floor, balancing playout continuity with global quality preservation.
    \item We evaluate \sys on a 16$\times$H100 GPU cluster. At comparable generation quality, \sys improves \textit{Quality of Experience} (QoE), measured by \textit{Continuous Play Ratio} (CPR), by 1.64$\times$--3.29$\times$ and reduces \textit{Time to First Chunk} (TTFC) by 1.61$\times$--9.65$\times$ over baselines.
\end{itemize}

\section{Background and Motivation}

Table~\ref{tab:terminology} summarizes the terminology used in the paper.

\subsection{Video Generation}
\label{sec:backvideogen}

As shown in Figure~\ref{fig:p1}(a), video generation~\cite{stablediffusion,peebles2023scalable,wan,hunyuanvideo,selfforcing,causalforcing} typically follows a three-stage latent \textit{diffusion pipeline}~\cite{ddim,ddpm,blattmann2023stable}. First, an encoder~\cite{t5xxl} maps conditional inputs, such as prompts and references, into conditioning embeddings. Second, conditioned on these embeddings, a DiT model iteratively denoises latent noise over multiple steps~\cite{ddim,ddpm} to generate a latent video. Third, a decoder~\cite{stablediffusion} maps the latent video into pixel-space video. Existing models built around this pipeline mainly fall into two paradigms: offline one-shot generation and chunk-wise autoregressive generation. 

\textbf{Offline DiT.}
As shown in Figure~\ref{fig:p1}(b), offline DiTs~\cite{peebles2023scalable} treat an entire video as a single diffusion process: all latent frames are denoised together to produce the whole video. Because all frames are processed jointly, attention spans the full video, causing attention cost to grow quadratically with the number of frames~\cite{transformer}. It therefore does not scale well to long videos or low-latency real-time settings~\cite{adaspa,svg}.

\textbf{AR-DiT.} AR-DiTs~\cite{selfforcing,causalforcing,rollingforcing,magi} reformulate video generation as a chunk-by-chunk process. As shown in Figure~\ref{fig:p1}(c), the model generates one short video segment at a time, which we call a \textit{chunk}. The chunk becomes immediately available for playout once generated and is then appended to a rolling KV cache for subsequent chunks.

To keep the KV cache bounded for long streams, AR-DiTs commonly adopt a \textit{sink + local} strategy~\cite{selfforcing,rollingsink,streamingllm}, which preserves the first few chunks as a global sink and keeps KV entries only for the most recent $W$ chunks. This allows a stream to extend, in principle, indefinitely under bounded GPU memory. CausVid~\cite{causvid}, Self-Forcing~\cite{selfforcing}, and Rolling Forcing~\cite{rollingforcing} follow this strategy, making AR-DiTs a key foundation for real-time streaming video generation.

\textbf{Fidelity knobs and configurations.}
Real-time video generation exposes several serving-time knobs that trade generation latency for visual fidelity. We call each such dimension a \textit{fidelity knob}, and a concrete assignment of these knobs a \textit{fidelity configuration}. In this paper, a fidelity configuration includes denoising steps $S$, attention sparsity $\rho$, KV-window size $W$, and quantization mode $Q$. (1) \textit{Denoising Steps.} This knob controls the number of denoising steps $S$ in DiT models. Reducing $S$ roughly lowers latency linearly, but can alter the speed--quality tradeoff in diffusion deployment~\cite{ddim,lu2025dpm,du2025fewer}. (2) \textit{Sparse Attention.} It skips non-selected attention blocks with sparse masks and is controlled by attention sparsity $\rho$~\cite{svg,sparsevdit,adaspa}. (3) \textit{KV Window Size.} It controls the maximum KV-cache window size $W$ used by attention, following sliding-window and attention-sink mechanisms in streaming autoregressive generation~\cite{streamingllm,rollingforcing,rollingsink}. (4) \textit{Quantization.} It compresses weights or activations into a lower-precision mode $Q$, such as FP8, exploiting tensor cores to accelerate compute and memory access~\cite{sageattention,flashattention3}. These knobs define the fidelity-configuration space for serving-time speed--quality tradeoffs in real-time video generation. Importantly, for a fixed target resolution, chunk size, and fidelity configuration, DiT chunk latency is highly profileable offline~\cite{tridentserve,tensorrtllm,diffserve}, providing the timing prior used by \sys for resource allocation (\S\ref{sec:resource}) and fidelity selection (\S\ref{sec:precision}).

\subsection{Resource Allocation in Serving Systems}
Serving systems allocate compute resources along two complementary dimensions. 

The first is temporal allocation, which determines which request should receive service earlier. Its central mechanism is preemption, where execution is reordered at safe boundaries to favor more urgent requests. Preemption can occur within a request or across requests. Within a request, techniques such as continuous batching~\cite{orca} expose fine-grained execution boundaries and allow the system to interleave work across iterations~\cite{vllm,sarathi-serve}. Across requests, priority scheduling preempts less urgent requests and lets latency-sensitive ones run earlier~\cite{distserve}. Temporal allocation changes when a request receives service, but not the amount of compute assigned to it at a given time.

The second dimension is spatial allocation, which changes a request's compute share by scaling execution across devices. Its central mechanism is parallelism, including tensor~\cite{tp}, pipeline~\cite{pp}, and sequence parallelism~\cite{ulysses,ring,loongserve,tridentserve}. For DiTs, sequence parallelism (SP) is especially relevant because video generation creates long token sequences. SP partitions the sequence across GPUs, lets each GPU process one shard, and uses collective communication to exchange states required by attention. Increasing the sequence-parallel degree can reduce per-chunk latency, but it also occupies more workers and may require state redistribution.

\subsection{Motivation}
\label{subsec:motivation}

\textbf{Limitations of existing serving systems.} For scheduling and resource allocation, existing LLM serving systems such as vLLM~\cite{vllm}, Sarathi-Serve~\cite{sarathi-serve}, DistServe~\cite{distserve}, and LoongServe~\cite{loongserve} are largely designed around request-level latency and throughput objectives, typically expressed as \textit{Time to First Token} (TTFT), \textit{Time per Output Token} (TPOT), or goodput~\cite{distserve}.
Text streaming serving systems, such as Andes~\cite{liu2024andes}, TokenFlow~\cite{chen2026tokenflow}, have explored satisfying user quality of experience (QoE) by introducing preemptive request scheduling methods. Although these systems draw inspiration from QoE in video streaming~\cite{balachandran2012quest}, the technical challenges are largely different and cannot be directly applied to real-time video generation serving (\S\ref{sec:relatedwork}).
Systems for diffusion and video serving, such as DiffServe~\cite{diffserve}, TridentServe~\cite{tridentserve}, and TetriServe~\cite{tetriserve}, mainly optimize throughput or completion time for offline generation workloads~\cite{tridentserve}. More recently, StreamDiffusionV2~\cite{streamdiffusionv2} targets streaming video generation through batching and pipeline parallelism~\cite{gpipe,pp}, but still optimizes frames-per-second (\textit{FPS}) objectives rather than explicitly tracking each stream's real-time playout slack. For acceleration knobs, methods such as AWQ~\cite{awq}, SparseVideoGen~\cite{svg}, and SageAttention~\cite{sageattention} improve speed through static quality-affecting configurations, which are chosen before serving and remain unchanged at runtime.

However, in real-time generation, the goal is not merely faster computation but preserving \textit{playout continuity}, which is critical to user experience. Existing serving systems do not explicitly exploit this objective.

\textbf{Motivating Insights.} With \textit{playout continuity} as the primary goal, we derive two system insights.

\underline{\textit{Insight 1.}} \textit{Compute resources should be reallocated according to playout slack.} Because playout slack evolves over time, preserving \textit{playout continuity} requires resource allocation to track this evolution and redirect compute toward streams under tighter playout pressure. Existing serving systems do not make allocation decisions based on this per-stream slack signal, so urgent streams can remain under-served and eventually stall. Our case study with StreamDiffusionV2 (\S\ref{subsec:casestudy}) shows that even when average system FPS matches the playout rate, some streams still stall due to the mismatch between playout slack and resource allocation.

\underline{\textit{Insight 2.}} \textit{Each chunk's fidelity configuration should adapt to the current playout slack.} Since users are highly sensitive to stalls, delivering an acceptable chunk on time is often more valuable than a late high-fidelity chunk. This principle is widely adopted in streaming media~\cite{bba,pensieve,mpc}. Offline video generation fixes one fidelity configuration for the whole video due to one-shot execution. In contrast, AR-DiTs generate one chunk at a time, allowing the system to reduce latency when slack is tight and restore fidelity when slack recovers (\S\ref{subsec:mainresult},~\S\ref{subsec:casestudy}).

These insights define the design space of \sys: 
Driven by playout slack, it reallocates resources across streams toward urgent streams (\S\ref{sec:resource}) and selects per-chunk fidelity configurations within streams to balance \textit{playout continuity} and visual quality (\S\ref{sec:precision}).

\begin{table}[t]
\centering
\caption{Terminology used in the problem formulation.}
\label{tab:terminology}
\small
\setlength{\tabcolsep}{4pt}
\renewcommand{\arraystretch}{0.95}
\begin{tabularx}{\columnwidth}{@{}>{\RaggedRight\arraybackslash}p{0.20\columnwidth}Y@{}}
\toprule
\textbf{Term} & \textbf{Meaning} \\
\midrule
\textit{Stream}
& One real-time video generation session. \\
\textit{Frame}
& A single image in the generated video stream. \\
\textit{Chunk}
& A short video segment containing multiple consecutive frames, generated and delivered as one unit in streaming video generation. \\
\textit{Playout continuity}
& Chunks are ready before playback reaches them, avoiding visible stalls. \\
\textit{Playout slack}
& Playable video buffer already generated but not yet consumed by playback. \\
\bottomrule
\end{tabularx}
\end{table}

\section{System Overview}
\label{sec:overview}

Based on these insights, we develop \sys, to our knowledge, the first playout-slack-driven serving system for real-time streaming video generation. 

Below, we first introduce the core runtime concepts of \sys (\S\ref{sec:core-concepts}), then its system components (\S\ref{sec:three-planes}), and finally its complete workflow (\S\ref{sec:workflow}).

\subsection{Core Runtime Concepts}
\label{sec:core-concepts}

\textbf{Service Credit.} \textit{Service credit} is derived from \textit{playout slack} to quantify stream urgency. Both resource-allocation and fidelity-selection decisions use \textit{service credit} as the primary control signal.
For each active stream $u$, we define its \textit{service credit} $C_u$ as:
\begin{equation}
C_u = P_u - (R_u + T_u)
\end{equation}
where $P_u$ is the playout slack (i.e., the remaining playable buffer), $R_u$ is the estimated remaining time to finish the current chunk (zero if the stream is not running), and $T_u$ is the profiled generation time of the next chunk under the fidelity configuration selected for the next chunk (\S\ref{sec:precision}). Thus, service credit converts the dynamic playout SLO into a per-stream urgency score.

\textbf{Step/Chunk Boundaries.} As shown in Figure~\ref{fig:p1}, a step boundary occurs after one DiT denoising step, while a chunk boundary occurs after one chunk completes. These boundaries are safe points for resource allocation and fidelity-configuration selection.

\textbf{Home Worker.} Each stream is assigned a \textit{home worker}\footnote{We use \emph{worker} uniformly for the unit of scheduling and execution. By default, one worker corresponds to one GPU and one model replica.} at admission. By default, a stream's generation and KV cache remain on its \textit{home worker}.

\begin{figure}[t]
\centering
\includegraphics[width=\linewidth]{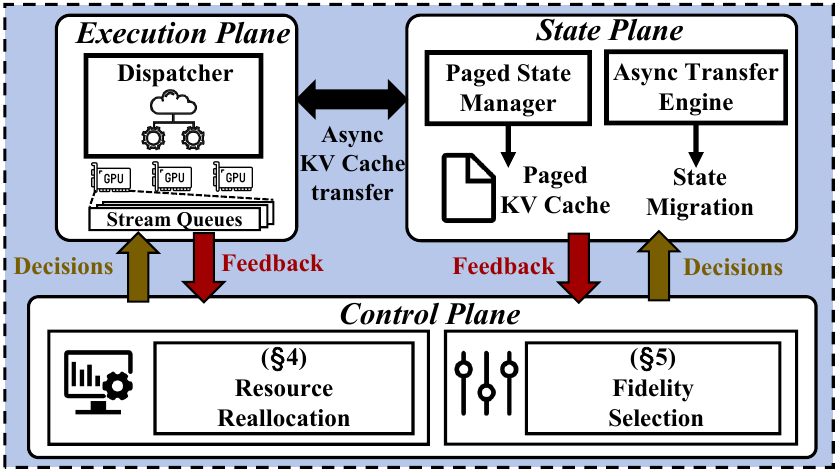}
\caption{System overview and components of \sys.}
\label{fig:p4}
\end{figure}

\subsection{System Components}
\label{sec:three-planes}

Figure~\ref{fig:p4} shows the system architecture. \sys is organized into three planes:

\begin{figure*}[t]
\centering
\includegraphics[width=0.9\linewidth]{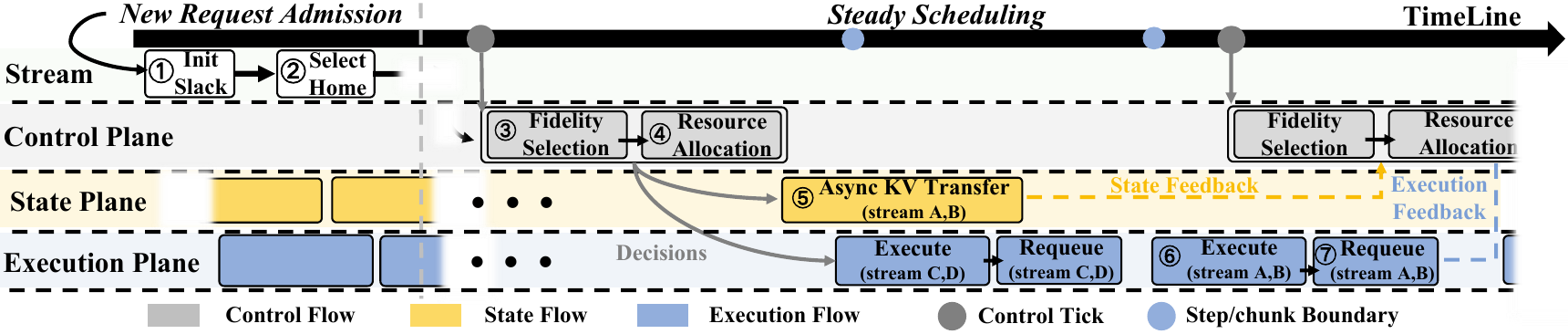}
\vspace{-0.15in}
\caption{System workflow of \sys.}
\label{fig:p5}
\end{figure*}

\textbf{Control Plane.} The \textit{Control Plane} is a global controller that wakes up at each \textit{control tick} (every \SI{3}{s} by default) to make resource-allocation and fidelity-selection decisions. It integrates \textit{Slack-Driven Resource Reallocation} (\S\ref{sec:resource}) and \textit{Slack-Driven Fidelity Selection} (\S\ref{sec:precision}) into a single control loop. It tracks each stream's service credit $C_u$, decides which streams should run at the next step or chunk boundary, and makes the corresponding resource-allocation and fidelity-configuration decisions to better preserve \textit{playout continuity}.

\textbf{State Plane.} The \textit{State Plane} (\S\ref{sec:stateplane}) uses a \textit{Paged State Manager} to manage KV cache and page GPU-resident KV at latent-frame granularity\footnote{Since different streams may have different KV lengths, frame-level paging reduces fragmentation.}. To support resource reallocation, it also includes an \textit{Async Transfer Engine} that asynchronously migrates KV state.

\textbf{Execution Plane.} The \textit{Execution Plane} maintains stream queues per worker. At each step or chunk boundary, the \textit{Execution Plane} dispatches the next stream according to decisions from the \textit{Control Plane}.

The three planes interact through lightweight interfaces, where the \textit{Control Plane} makes \textit{decisions} to guide the \textit{Execution} and \textit{State Planes}, while the \textit{Execution Plane} and \textit{State Plane} provide feedback for future decisions.

\subsection{System Workflow}
\label{sec:workflow}

We now describe the workflow enabled by the three-plane design in Figure~\ref{fig:p5}.
For each stream, the workflow consists of two phases, \textit{stream admission} and \textit{steady scheduling}.

\textbf{Stream admission.}
When a new request arrives, \textcircled{1} \sys assigns an initial playout slack as its \textit{Time to First Chunk} (TTFC) constraint, set to $4\times$ the estimated first-chunk generation time by default.
\textcircled{2} \sys then selects the worker with the fewest streams as the stream's \textit{home worker}, avoiding already congested workers. After admission, the stream enters \textit{steady scheduling}.

\textbf{Steady scheduling.}
In steady scheduling, the system is driven by two types of events, periodic control ticks and streams reaching step or chunk boundaries.

At each control tick, the \textit{Control Plane} makes execution and migration decisions for the next step or chunk boundary. \textcircled{3} First, based on current playout slack, it performs fidelity selection (\S\ref{sec:precision}) and chooses each stream's next-chunk fidelity configuration. \textcircled{4} Second, it performs resource reallocation (\S\ref{sec:resource}) and decides which streams should run and where they should run. Given the selected fidelity configurations, it recomputes stream service credits to determine which streams should run at the next step or chunk boundary (\S\ref{sec:queue}) and on which worker(s) they should run (\S\ref{sec:rehome} and \S\ref{sec:elastic}). These joint decisions shift resources across streams toward urgent streams to better preserve \textit{playout continuity}. Third, if a decision requires state migration, \textcircled{5} the \textit{State Plane} (\S\ref{sec:stateplane}) starts asynchronous KV transfer and overlaps it with computation when possible. 

At each step or chunk boundary, \textcircled{6} the \textit{Execution Plane} dispatches and executes streams using the assigned workers and fidelity configurations according to the \textit{Control Plane}'s decisions. \textcircled{7} After each step, streams are requeued for the next dispatch decision. Unless the \textit{Control Plane} updates its decision at a later control tick, streams continue under the current decision.

In \S\ref{sec:resource} and \S\ref{sec:precision}, we detail how the \textit{Control Plane} guides resource reallocation and fidelity selection, respectively. Appendix~\ref{app:control-workflow} summarizes the complete control loop.

\section{Slack-Driven Resource Reallocation}
\label{sec:resource}
This section details how \sys reallocates compute resources in real-time video generation.

In a nutshell, our goal is to redirect compute resources from streams with sufficient playout slack to streams at risk of stalling. 
To achieve this goal, there are two general strategies.
The first is preemptive scheduling, which prioritizes urgent streams and defers non-urgent ones. The second is per-stream compute expansion, which increases parallelism by borrowing compute resources for an urgent stream so it can finish before its playout slack is exhausted.

As shown in Table~\ref{tab:mech}, \sys realizes these strategies with three mechanisms: per-worker three-tier priority queuing for preemption on each worker (\S\ref{sec:queue}), cross-worker re-homing for redistributing streams across workers (\S\ref{sec:rehome}), and parallelism scaling for temporary compute expansion (\S\ref{sec:elastic}). Finally, \S\ref{sec:stateplane} presents the State Plane that supports state migration required by these mechanisms.

Guided by service credit, the \textit{Control Plane} composes these mechanisms into a staged resource-allocation loop. First, it orders streams in each worker's queue, enabling local preemption at the next boundary. Then, it detects cluster-wide imbalance and generates a re-homing plan for cross-worker preemption. Finally, for streams still prone to violate \textit{playout continuity}, it expands computing through parallelism scaling. These approaches can be combined under severe slack pressure.

\begin{table}[t]
\centering
\caption{Resource reallocation mechanisms, organized by preemption and compute expansion.}
\vspace{-0.1in}
\label{tab:mech}
\small
\setlength{\tabcolsep}{4pt}
\renewcommand{\arraystretch}{1.15}
\begin{tabularx}{\columnwidth}{
  >{\bfseries\RaggedRight\arraybackslash}p{0.29\columnwidth}
  >{\RaggedRight\arraybackslash}X
  >{\RaggedRight\arraybackslash}X
}
\toprule
 & \textbf{Expansion (\xmark)} & \textbf{Expansion (\cmark)} \\
\midrule
Preemption (\xmark)
& --
& Elastic SP \newline (\S\ref{sec:elastic}) \\
\addlinespace[2pt]
Preemption (\cmark)
& Three-tier queue \newline (\S\ref{sec:queue}), \newline
  re-homing \newline (\S\ref{sec:rehome})
& Three-tier queue, \newline
  re-homing, \newline
  elastic SP \newline
  (\S\ref{sec:queue}--\ref{sec:elastic}) \\
\bottomrule
\end{tabularx}
\end{table}

\begin{figure*}[t]
\centering

\begin{minipage}[t]{0.595\textwidth}
\centering
\includegraphics[width=\linewidth]{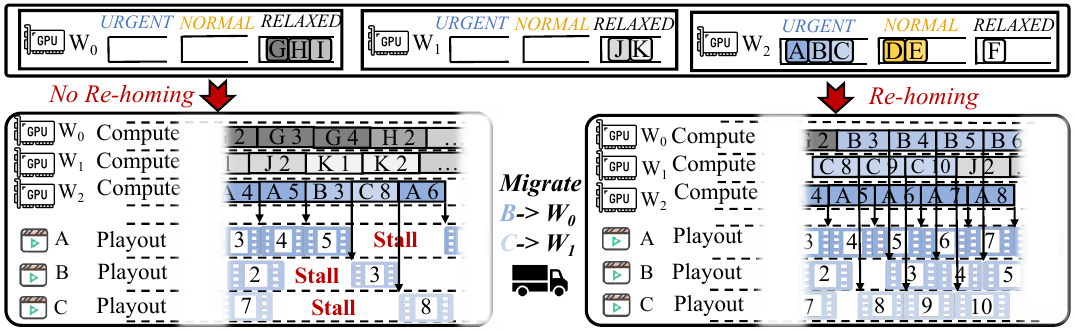}
\caption{Comparison with and without re-homing.}
\label{fig:p7}
\end{minipage}
\hfill
\begin{minipage}[t]{0.40\textwidth}
\centering
\includegraphics[width=\linewidth]{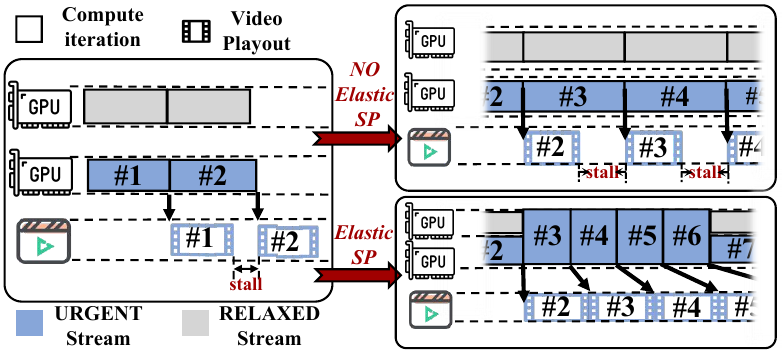}
\caption{Comparison with and without elastic SP.}
\label{fig:p8}
\end{minipage}

\end{figure*}

\subsection{Three-Tier Priority Queue and Preemption}
\label{sec:queue}

Three-tier priority queues provide local preemption and the urgency signal used by later reallocation mechanisms.

At each control tick, the \textit{Control Plane} computes each stream's service credit $C_u$ and classifies it as \textit{URGENT} ($C_u < 2T_u$), \textit{NORMAL} ($2T_u \le C_u \le 4T_u$), or \textit{RELAXED} ($C_u > 4T_u$). It then orders streams within each worker by service credit, so lower-credit streams are dispatched earlier at step or chunk boundaries. This local preemption prevents slack-deficient streams from waiting behind relaxed ones and supplies the tier information used by re-homing and elastic SP.

\begin{figure}[t]
\centering
\includegraphics[width=\linewidth]{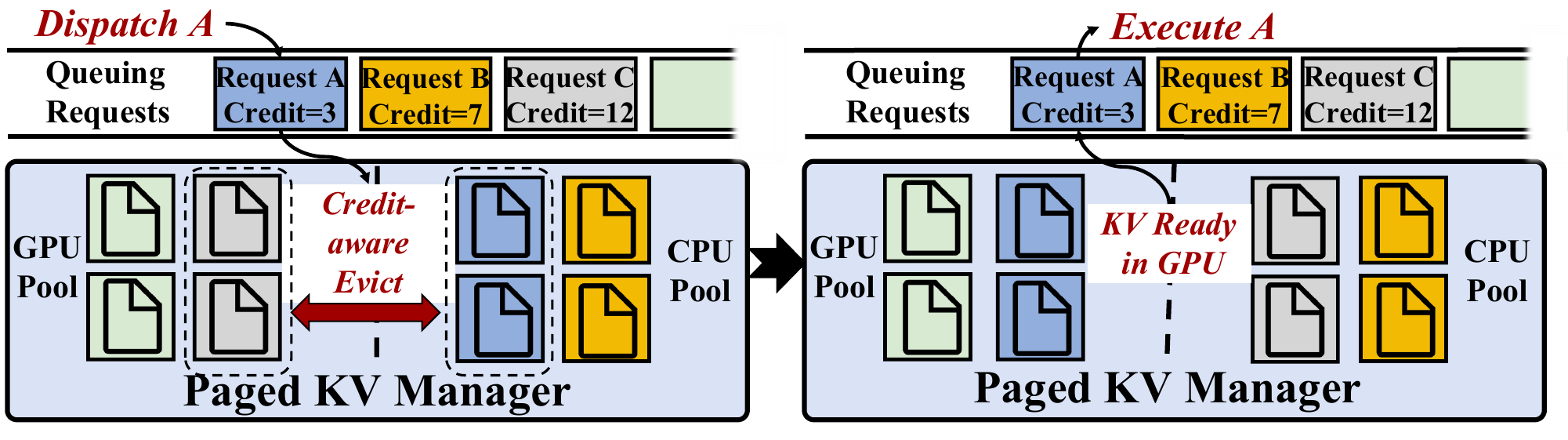}
\caption{Credit-aware eviction example.}
\label{fig:p6}
\end{figure}

\textbf{Credit-aware Eviction.}
Because preempted streams may leave their KV cache in a worker's limited GPU KV pool, the most urgent stream may find its state non-resident when it is rescheduled. When the pool is full, \sys evicts the highest-credit resident stream, i.e., the stream least likely to stall, to make room for the urgent stream's state (Figure~\ref{fig:p6}).

\subsection{Re-homing}
\label{sec:rehome}

Each stream is bound to a \textit{home worker} at admission and does not migrate by default, because it carries state such as KV cache, and frequent migration incurs substantial communication overhead.
In practice, however, we often observe an imbalance between URGENT-heavy workers and RELAXED-only workers (\S\ref{subsec:casestudy}). We refer to them as \textit{URGENT workers} and \textit{RELAXED workers}, respectively. Migrating URGENT streams from \textit{URGENT workers} to \textit{RELAXED workers} can relieve local congestion and improve workload balance. Figure~\ref{fig:p7} shows a typical case, where W0 and W1 have only RELAXED streams while W2 has several queued URGENT streams. Without migration, low-credit streams on W2 would repeatedly stall. \sys re-homes URGENT streams to W0 and W1, redirecting RELAXED-worker capacity toward urgent streams.

\textbf{Bipartite Re-homing Planning.}
We generate migrations using \textit{Bipartite Re-homing Planning}, which performs capacity-bounded, intra-node-preferred matching.

At each control tick, \sys matches overloaded workers to slack-rich workers. It treats workers with congested URGENT queues as senders and workers with no URGENT or NORMAL streams as receivers, so migration moves urgent work only toward spare slack capacity (lines~1--2). Receivers are tried in intra-node-first order to reduce transfer cost (line~5). Subject to send/receive caps, \sys moves the lowest-credit URGENT stream not in cooldown to an available receiver at its next chunk boundary (lines~6--8). This bounded plan improves balance while limiting migration bursts and repeated movement. Appendix~\ref{app:rehoming-details} gives the concrete details.

\begin{algorithm}[t]

\caption{Bipartite Re-homing Planning}
\label{alg:rehome}
\begin{algorithmic}[1]
\Require worker queues with urgency tiers and service credits
\Ensure migration plan $plan$
\State $senders \gets$ URGENT-heavy workers
\State $receivers \gets$ workers with no URGENT or NORMAL streams
\State $plan \gets \emptyset$
\For{$src \in senders$}
  \For{$dst \in receivers$ in intra-node-first order}
    \If{$src.sent < cap_{\mathrm{send}}$ and $dst.recv < cap_{\mathrm{recv}}$}
      \State $s \gets$ lowest-credit URGENT stream on $src$ not in cooldown
      \State append $(s,src,dst)$ to $plan$; update caps; mark $s$ in cooldown
    \EndIf
  \EndFor
\EndFor
\State \Return $plan$
\end{algorithmic}
\end{algorithm}

\subsection{Elastic Sequence Parallelism}
\label{sec:elastic}

Priority scheduling and re-homing may still fail to recover streams with extremely low service credit, due to estimation errors or runtime fluctuations. As shown in Figure~\ref{fig:p8}, \sys uses elastic SP as a last-resort recovery mechanism when a stream has $C_u < 0$, meaning it is projected to miss its next playout window. In this case, \sys borrows the highest-credit RELAXED worker as a donor and switches the stream to the corresponding pre-initialized intra-node SP2 group, accelerating the current chunk through sequence parallelism. The donor is released at the next safe boundary once the stream recovers to the NORMAL tier ($C_u \ge 2T_u$).

By default, \sys restricts elastic SP to intra-node SP2 and borrows at most one donor worker. All candidate SP2 groups are pre-initialized before serving, so elastic SP only switches the stream's active execution group rather than creating communication groups on the critical path. The required head-partition KV transfer is handled asynchronously by the State Plane (\S\ref{sec:stateplane}); Appendix~\ref{app:elastic-sp-details} and Appendix~\ref{app:sp-kv-transfer} give the full policy and transfer details.

\subsection{State Plane}
\label{sec:stateplane}

Credit-aware eviction (\S\ref{sec:queue}), re-homing (\S\ref{sec:rehome}), and elastic SP (\S\ref{sec:elastic}) are functionally distinct but share the same KV-state transfer primitive. \sys therefore centralizes state movement in an independent \textit{State Plane}, with a single transfer interface, shared non-blocking layer-wise streaming, and atomic-safety guarantees.

\begin{figure}[t]
\centering
\includegraphics[width=\linewidth]{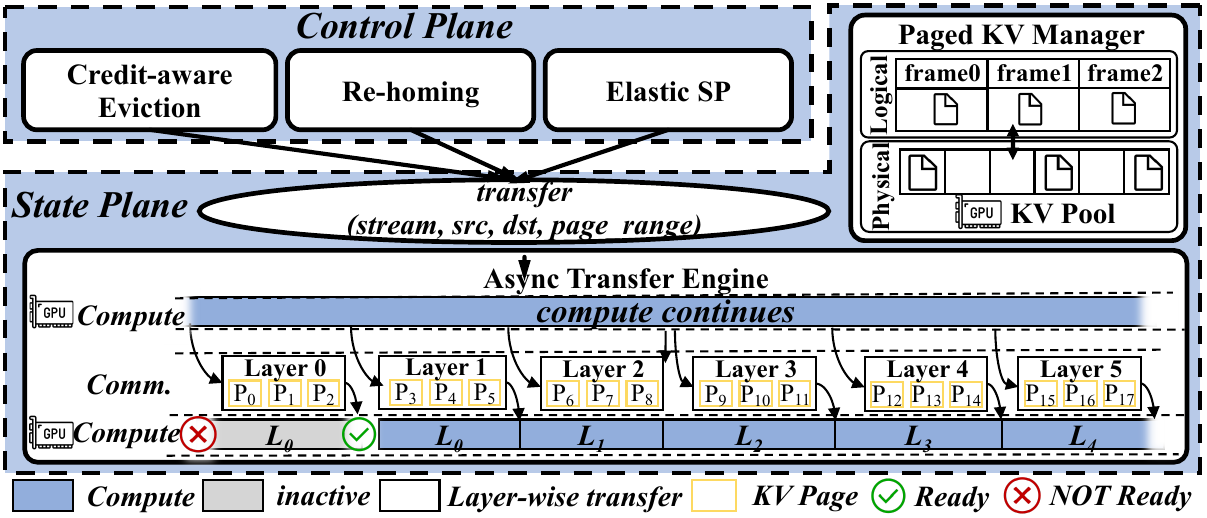}
\caption{State Plane design with unified KV management and asynchronous streaming transfer.}
\label{fig:p9}
\end{figure}

\textbf{Unified KV management.}
As shown in Figure~\ref{fig:p9}, each worker maintains a paged KV pool, defaulting to $\kappa=0.8$ of available VRAM~\cite{vllm}. Pages are allocated at frame granularity (one page per latent frame) and mapped to physical pages through a logical page table, avoiding fragmentation.

The three state-transfer operations share one interface:
\begin{verbatim}
transfer(stream, src, dst, page_range)
\end{verbatim}
where each caller specifies the required \texttt{page\_range}. This lets mechanisms in \S\ref{sec:queue}--\S\ref{sec:elastic} specify only which pages should reside where, leaving transfer timing to the \textit{State Plane}.

\textbf{Asynchronous streaming transfer.}
Synchronous KV movement can block dispatch and erase the benefit of reallocation. In \sys, transfer requests return immediately and are executed by an \textit{Async Transfer Engine} on dedicated CUDA streams.
The engine uses one protocol for eviction, re-homing, and elastic SP (Figure~\ref{fig:p9}):
\begin{itemize}
    \item \textit{Non-blocking}. Page transfers are scheduled asynchronously subject to dependency and bandwidth limits, so foreground dispatch and chunk computation do not wait for a full transfer to complete.
    \item \textit{Atomic safety}. The dispatcher must never schedule a stream with an incomplete state. Once a transfer is submitted, the target stream is temporarily removed from the active queue. The stream is reinserted once its first-layer state is ready, enabling layer-wise computation transfer overlap while preserving correctness.
    \item \textit{Layer-wise streaming}. 
    Transfers are issued in layer-major order. Once layer $k$'s pages arrive, the destination can start layer-$k$ computation while later layers continue transferring, overlapping computation with state movement.
\end{itemize}

\section{Slack-Driven Fidelity Selection}
\label{sec:precision}

This section introduces how \sys makes per-chunk fidelity-configuration decisions, applied at chunk boundaries, based on playout slack. We construct a fidelity-configuration space from four knobs: denoising steps $S$, attention sparsity $\rho$, KV-window size $W$, and quantization mode $Q$ (detailed in Appendix~\ref{app:profile}). A concrete assignment forms a fidelity configuration $\mathrm{cfg}_i=\{S_i,\rho_i,W_i,Q_i\}$. These knobs control the speed--quality tradeoff\footnote{We use VBench~\cite{vbench}, a widely used video-generation benchmark, to measure generation quality.}.

Making fidelity-selection decisions is challenging. On the one hand, the complex multidimensional search space makes online fidelity-configuration selection difficult under a playout-slack budget. On the other hand, choosing only the fastest fidelity configuration within the playout-slack budget can push the system toward low-quality fidelity configurations under overload, causing visible degradation.

To address this, we propose \textit{Bi-Modal Pareto Routing (BMPR)}. BMPR uses an empirical latency--quality Pareto frontier and a global quality floor to make a bi-modal decision: it prioritizes quality when playout slack permits, and enforces the quality floor when playout slack is tight.

We first present a key observation (\S\ref{sec:obs}) that motivates our design. We then describe BMPR's offline construction and online selection procedure (\S\ref{sec:bmpr}).

\subsection{Observation: quality loss has limited propagation through KV cache}
\label{sec:obs}

\begin{figure}[t]
\centering
\includegraphics[width=\linewidth]{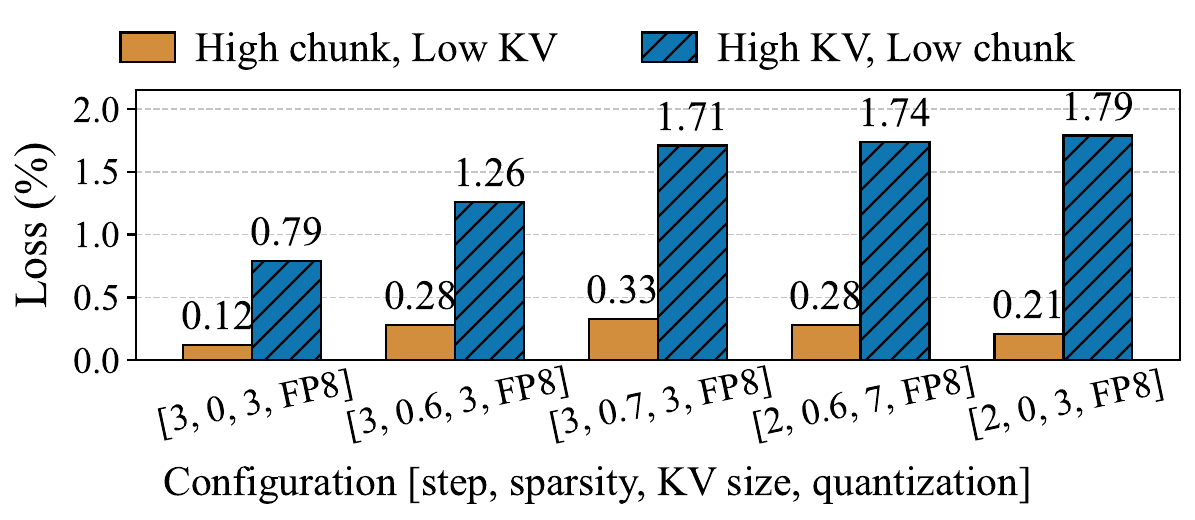}
\caption{Quality impact of low-precision current chunks and low-precision historical KV cache.}
\label{fig:p10}
\end{figure}

In Figure~\ref{fig:p10}, we test whether a low-precision configuration used for previous chunks degrades the current chunk through the rolling KV cache. 

We isolate the source of quality loss by comparing two settings against the all-high-quality reference. In the first setting, the current chunk uses the highest-quality fidelity configuration, while its KV cache is produced by a low-cost fidelity configuration $\mathrm{cfg}_i$. In the second setting, the current chunk uses $\mathrm{cfg}_i$, while its KV cache is produced by the highest-quality fidelity configuration. We measure the percentage drop in VBench~\cite{vbench} for the current chunk. The results show that low-fidelity historical KV causes only a small quality drop, whereas lowering the current chunk's fidelity configuration causes a much larger drop.

This observation shows that quality degradation has limited propagation through KV cache, making per-chunk fidelity-configuration decisions largely independent: downshifting an urgent chunk does not imply persistent quality degradation in later chunks.

\subsection{Bi-Modal Pareto Routing}
\label{sec:bmpr}

Based on this observation, we present Bi-Modal Pareto Routing (BMPR), which uses an offline empirical latency--quality Pareto frontier with a global quality floor $Q_{\text{floor}}$ for online fidelity selection.

\textbf{Offline construction.}
First, \sys profiles each candidate's fidelity configuration and represents it as a point $(L,Q)$ in the latency--quality plane, with details given in Appendix~\ref{app:profile}. A fidelity configuration $i$ is dominated if another fidelity configuration $j$ satisfies $L_j \le L_i$ and $Q_j \ge Q_i$, with at least one strict inequality. The non-dominated fidelity configurations form the empirical Pareto frontier:
\[
F = \big\{(L_i, Q_i, \text{cfg}_i) \mid \text{cfg}_i \text{ is non-dominated}\big\}.
\]
This frontier removes fidelity configurations for which another fidelity configuration is no slower and no lower-quality, so online selection considers only meaningful speed--quality tradeoff points.

Second, to avoid visible degradation from aggressive downshifting, \sys uses the median quality over all candidate fidelity configurations as the global quality floor $Q_{\text{floor}}$.

\textbf{Online selection.}
Given a playout-slack budget $B$ for the current chunk, BMPR selects a fidelity configuration from $F$ in two modes:
\begin{itemize}
    \item \textit{Quality mode.} By default, BMPR filters frontier points with $L_i \le B$ and $Q_i \ge Q_{\text{floor}}$. If the set is non-empty, BMPR selects the highest-quality fidelity configuration. This mode targets normal or mildly congested conditions, preserving quality while staying within the playout-slack budget.
    \item \textit{Speed-recovery mode.} Quality mode is infeasible when no configuration within the budget also satisfies the quality floor. In this case, BMPR avoids blindly choosing the fastest point regardless of quality. Instead, it selects the min-latency point among fidelity configurations with $Q_i \ge Q_{\text{floor}}$. This choice may exceed the current budget but prevents visible quality loss.
\end{itemize}

The speed-recovery mode may still violate \textit{playout continuity}. In that case, the resource reallocation mechanisms in \S\ref{sec:resource} provide the next line of defense. This joint design of bi-modal fidelity selection and resource allocation preserves \textit{playout continuity} while bounding visual degradation. \S\ref{subsec:sensitivity} shows that BMPR maintains stable quality even when higher arrival rates create more low-credit chunks, rather than degrading linearly with congestion.

\section{Implementation}

\sys is implemented in a custom serving framework of about 18K lines of Python code and supports representative AR-DiT models, including CausVid~\cite{causvid}, Self-Forcing~\cite{selfforcing}, Causal-Forcing~\cite{causalforcing}, and Rolling Forcing~\cite{rollingforcing}. We use Ray to manage cluster workers~\cite{ray}. The \textit{Control Plane} runs on an \texttt{asyncio} event loop~\cite{asyncio} and coordinates through Ray actors~\cite{ray}.
\sys also includes an offline profiler that provides latency and quality profiles for fidelity selection and service-credit estimation.

For fidelity configurations, we implement sparse attention based on Light Forcing~\cite{lightforcing} and use SageAttention2~\cite{sageattention2} for online FP8 attention quantization. SageAttention2 dynamically quantizes and dequantizes activations online without quantizing model weights, so switching fidelity configurations does not require weight reloading~\cite{sageattention2}.

The unified transfer interface in \S\ref{sec:stateplane} is implemented by a NIXL-based transfer engine~\cite{nixl}. It uses NCCL P2P over NVLink within a node~\cite{nccl,nvlink} and the NIXL RDMA backend across nodes~\cite{nixl,gpudirectrdma}. All transfers run on independent CUDA streams and synchronize with compute streams through CUDA events~\cite{cuda}.

\begin{figure*}[t]
\centering
\includegraphics[width=\linewidth]{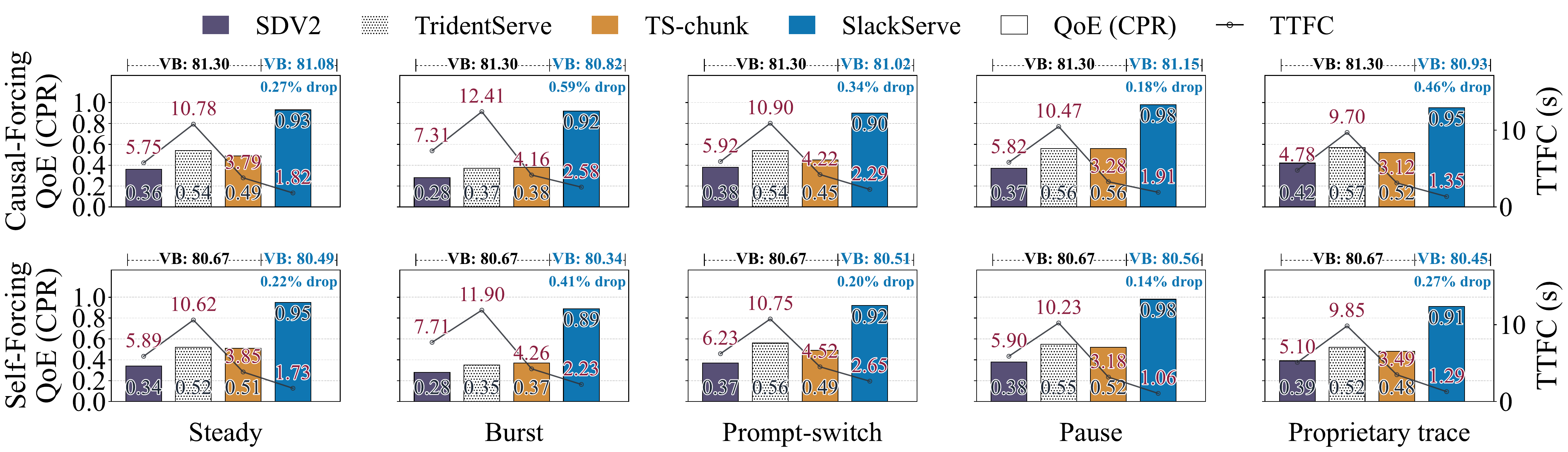}
\vspace{-0.25in}
\caption{End-to-end comparison on Causal-Forcing and Self-Forcing across the five workloads.}
\label{fig:p12}
\end{figure*}

\section{Evaluation}

\subsection{Experimental setup}

\textbf{Testbed.} We evaluate \sys on a two-node cluster with 16 NVIDIA H100 80GB GPUs. GPUs within a node are connected by NVLink with up to 900~GB/s per GPU~\cite{h100,nvlink}, and the two nodes are connected by 400~Gb/s InfiniBand~\cite{quantum2}.

\textbf{Models.} We use two representative AR-DiT models, Causal-Forcing~\cite{causalforcing} and Self-Forcing~\cite{selfforcing}. By default, each chunk contains three latent frames, with 480p resolution and a 16~fps playout rate.

\textbf{Workloads.} We construct five workloads: four synthetic ones and one proprietary enterprise trace for evaluation.
\begin{itemize}
    \item \textbf{Steady}: streams arrive according to a Poisson process~\cite{poisson} at 1 stream/s.
    \item \textbf{Burst}: extends Steady with three burst points, each causing 10\% of all streams to arrive simultaneously.
    \item \textbf{Prompt-switch}: extends Steady by injecting condition switches at random positions in every stream, resetting playout slack.
    \item \textbf{Pause}: extends Steady by adding client-side pauses to every stream, during which playout stops and slack accumulates.
    \item \textbf{Proprietary trace}: a scaled enterprise trace with mixed steady, bursty, and idle arrival patterns.
\end{itemize}
We use all 946 prompts from VBench~\cite{vbench} and sample each stream length from $\{81,129,161,241\}$ frames, corresponding to about 5--15~s. Appendix~\ref{app:workload} details all workloads.

\noindent\textbf{Baselines.}
We compare SlackServe against three representative external baselines.
\begin{itemize} 
    \item \textbf{StreamDiffusionV2 (SDV2)}~\cite{streamdiffusionv2} represents streaming-video serving that optimizes frame-rate objectives with FIFO and batching.
    \item \textbf{TridentServe (TS)}~\cite{tridentserve} represents DiT serving with dynamic resource management, including dynamic parallelism and migration, but manages SLOs at the per-stream level.
    \item \textbf{TridentServe-chunk (TS-chunk)} is a slack-aware baseline which extends TS to per-chunk SLO management: it uses least-slack-first scheduling with static fidelity, while retaining TS's dynamic parallelism and re-homing mechanisms.
\end{itemize}
Because no existing open-source system fully supports playout-slack-driven AR-DiT serving, \S\ref{subsec:ablation} uses an ablation study to evaluate stronger controlled variants of SlackServe that cover the main competing design points.

\textbf{Metrics.} We use three metrics that directly reflect the real-time video experience.
\begin{itemize}
    \item 
    \textbf{Quality of Experience (QoE)}: We define QoE as user-side playout continuity, measured by \textit{Continuous Play Ratio} (CPR), the fraction of chunks delivered before their playout deadlines: \[ \mathrm{QoE} \equiv \mathrm{CPR} = \frac{1}{|\mathcal{U}|}\sum_{u\in\mathcal{U}} \frac{1}{N_u}\sum_{i=1}^{N_u} \mathbf{1}\{t_{u,i}^{\mathrm{ready}} \le t_{u,i}^{\mathrm{ddl}}\}. \] Here $t_{u,i}^{\mathrm{ready}}$ is the ready time of chunk $i$ in stream $u$, and $t_{u,i}^{\mathrm{ddl}}$ is its playout deadline. Higher QoE indicates fewer playout-continuity violations.
    \item \textbf{Time to First Chunk (TTFC)}: the time from stream arrival to delivery of the first playable chunk to the client, analogous to time-to-first-output metrics in generative serving~\cite{distserve}.
    \item \textbf{VBench}~\cite{vbench}: a widely used composite benchmark for video generation, covering visual quality, temporal consistency, motion quality, and semantic alignment.
\end{itemize}

\subsection{End-to-end results}
\label{subsec:mainresult}

We compare \sys with StreamDiffusionV2 (SDV2), TridentServe (TS), and TridentServe-chunk (TS-chunk) across both models and all five workloads in Figure~\ref{fig:p12}.

\sys consistently improves QoE and reduces TTFC while maintaining comparable generation quality.
Across the five workloads and two models, \sys improves QoE by 2.65$\times$, 1.88$\times$, and 1.98$\times$ on average over SDV2, TS, and TS-chunk, respectively, and reduces TTFC by 3.39$\times$, 6.10$\times$, and 2.11$\times$. Across individual settings, these improvements span 1.64$\times$--3.29$\times$ for QoE and 1.61$\times$--9.65$\times$ for TTFC. The VBench drop is below 0.6\% across all settings, indicating comparable benchmark-level generation quality~\cite{vbench,svg,adaspa}.

The workload breakdown shows that \sys is effective under different pressure patterns. Under Steady and Burst, \sys keeps QoE high despite continuous load and synchronized arrivals. Under Prompt-switch, where condition changes reset accumulated slack, \sys maintains QoE of 0.90 on Causal-Forcing and 0.92 on Self-Forcing, compared with 0.37--0.56 for the baselines. Pause is less adversarial because client-side pauses accumulate slack, yet \sys again achieves the highest QoE, 0.98 on both models. On the proprietary trace, \sys sustains QoE of 0.95 on Causal-Forcing and 0.91 on Self-Forcing, showing that the same control policy remains effective under realistic non-stationary arrivals.

The baselines fall short for different reasons. SDV2 uses batching and pipeline parallelism to improve aggregate FPS, but this increases per-chunk latency and does not prioritize streams whose playout slack is being consumed. TS uses dynamic parallelism to accelerate requests and improves QoE over SDV2, but its control loop operates at the per-stream SLO level and cannot adapt individual chunks to the current slack budget. Its parallelism reconfiguration also delays the first chunk, inflating TTFC. TS-chunk is a slack-aware baseline that uses chunk-level SLOs to guide dynamic parallelism and scheduling, but its policy is designed for offline DiT serving. When applied to AR-DiT streaming, it frequently changes SP degree and migrates KV cache, introducing reconfiguration overhead that is unnecessary for many chunks.

\subsection{Ablation study}
\label{subsec:ablation}

We run ablations with Causal-Forcing on the Steady workload, varying one design dimension at a time.

\begin{figure}[t]
\centering
\includegraphics[width=\linewidth]{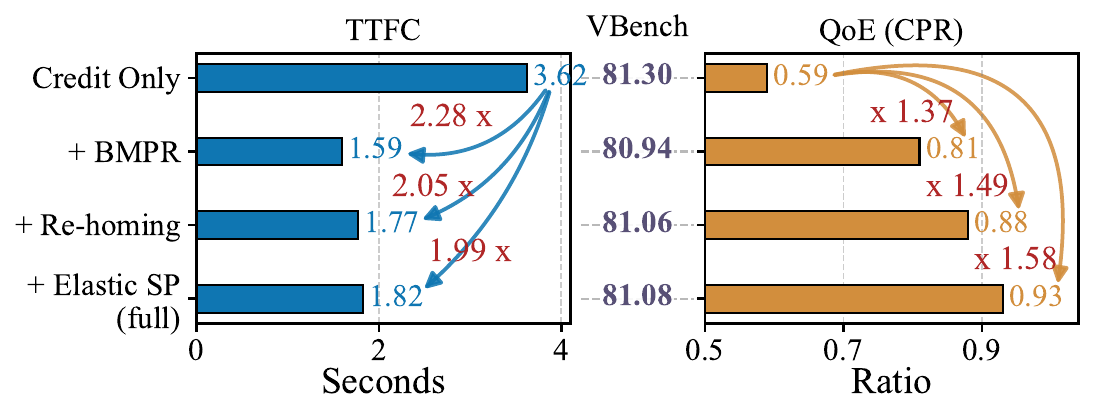}
\vspace{-0.2in}
\caption{Technique ablation with mechanisms enabled incrementally in Control-Plane trigger order.}
\label{fig:p13}
\end{figure}

\textbf{Technique ablation.}
We enable mechanisms in the order in which the \textit{Control Plane} invokes them (Figure~\ref{fig:p13}). In this ablation, \textit{Credit Only} corresponds to least-slack-first scheduling with static fidelity and without dynamic resource reallocation. Credit scheduling alone reaches 0.59 QoE, because boundary preemption gives low-credit streams earlier execution opportunities. Adding BMPR raises QoE to 0.81 by shortening urgent chunks, with only a 0.44\% quality drop. Re-homing provides the largest resource-side gain, improving QoE to 0.88 by moving URGENT streams to RELAXED workers. Elastic SP adds the final 0.05, reaching 0.93, by accelerating streams projected to stall even after local scheduling and re-homing. Quality changes little across the resource mechanisms, confirming that they primarily change \emph{where} resources are allocated rather than the fidelity configuration used by each chunk. TTFC first drops from 3.62s to 1.59s after BMPR, then slightly increases to 1.82s as re-homing and elastic SP add migration and coordination overhead.

\textbf{Comparison with fixed-level fidelity switching.}
We compare BMPR with a fixed-level switching policy that uses three Pareto-frontier configurations---fast, medium, and slow---and switches among them based on playout slack (Figure~\ref{fig:p14}). Although this policy reaches 0.86 QoE and 80.52 quality, BMPR further improves QoE to 0.93, increases quality by 0.70\%, and reduces TTFC from 2.50s to 1.82s. This shows that BMPR’s benefit does not come merely from dynamic fidelity switching, but from using the Pareto frontier and quality-floor constraint.

\begin{figure}[t]
\centering
\includegraphics[width=\linewidth]{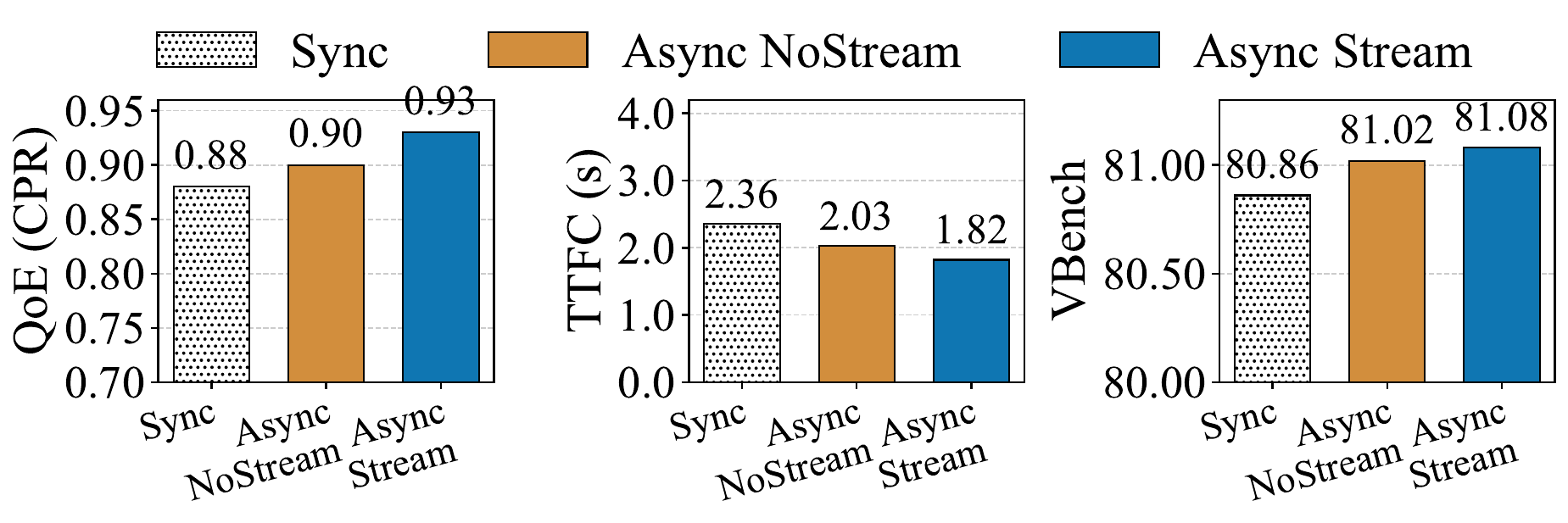}
\vspace{-0.2in}
\caption{Comparison of State-Plane transfer protocols.}
\label{fig:p15}
\end{figure}

\textbf{Comparison with different transfer protocols.}
Re-homing, elastic SP, and eviction all move the KV state. If these transfers block dispatch or computation, they can offset the scheduling benefit (Figure~\ref{fig:p15}). \textbf{Sync} blocks the dispatcher until migration, SP splitting, or eviction completes. \textbf{Async-NoStream} submits transfers asynchronously but waits for the entire state before target-side computation starts. \textbf{Async-Stream}, the default in \sys, adds layer-wise streaming and atomic readiness callbacks. Async-Stream reduces TTFC from 2.36s to 1.82s and improves QoE from 0.88 to 0.93. Quality also improves over Sync, since lower blocking overhead leaves more slack for BMPR to stay in quality mode.

\subsection{Case study}
\label{subsec:casestudy}

\begin{figure}[t]
\centering

\begin{minipage}[t]{0.46\linewidth}
\centering
\includegraphics[width=\linewidth]
{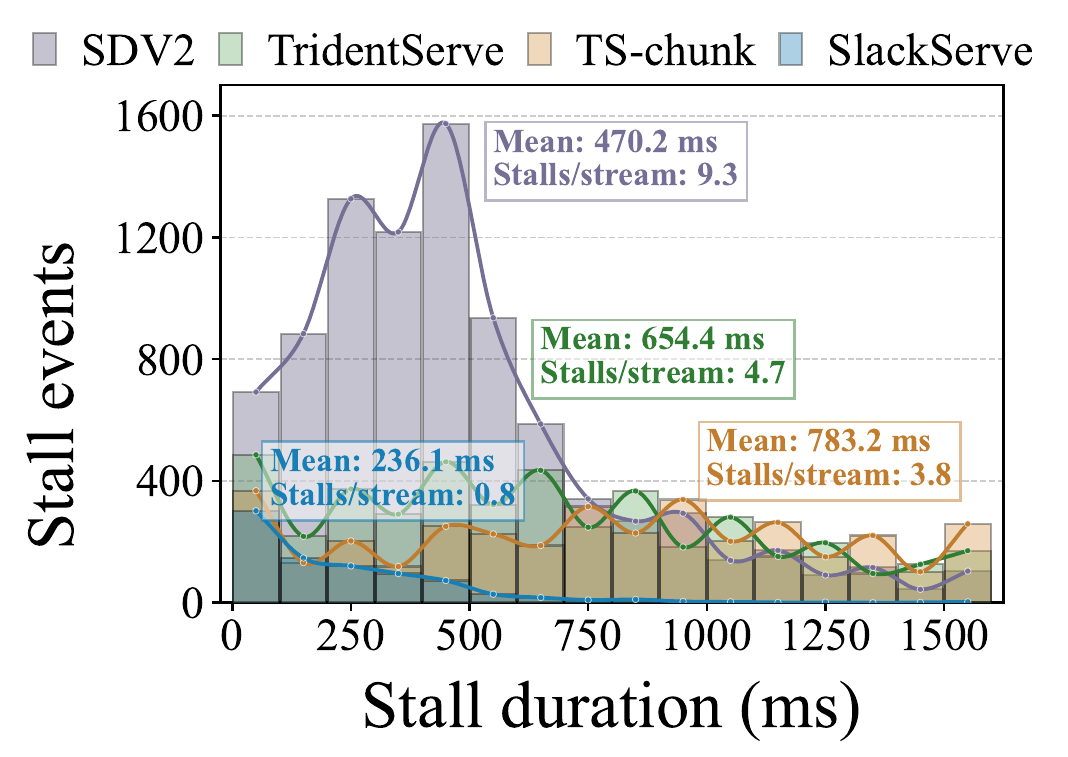}
\captionof{figure}{Stall-event duration distribution on the Steady workload.}
\vspace{-0.2in}
\label{fig:stall}
\end{minipage}
\hfill
\begin{minipage}[t]{0.46\linewidth}
\centering
\includegraphics[width=\linewidth]
{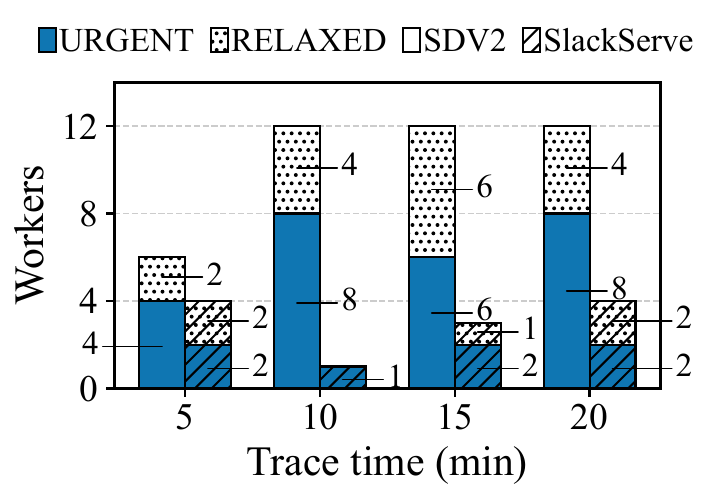}
\captionof{figure}{Worker-type distributions for \sys and SDV2.}
\vspace{-0.2in}
\label{fig:p16}
\end{minipage}
\end{figure}

\noindent\textbf{Stall analysis.}
Figure~\ref{fig:stall} shows the stall duration distribution on Steady across systems. \sys reduces average stall time from 470--783 ms to 236 ms, and stall frequency from 3.8--9.3 to 0.8 stalls per stream. This confirms that the QoE improvement in Figure~\ref{fig:p12} corresponds to fewer and shorter visible stalls, not merely a higher deadline-hit ratio.

\noindent\textbf{Imbalance analysis.}
Figure~\ref{fig:p16} explains why aggregate FPS alone is insufficient. Although SDV2's average FPS (16.8) reaches the playout target, it leaves 6.5 URGENT and 4.5 RELAXED workers on average, so urgent streams stall while relaxed workers still hold slack. SlackServe keeps URGENT and RELAXED workers near 1.75 and 1.25 through re-homing and local prioritization, which explains the stall reduction.

\begin{figure}[t]
\begin{minipage}[t]{0.46\linewidth}
\centering
\includegraphics[width=\linewidth]
{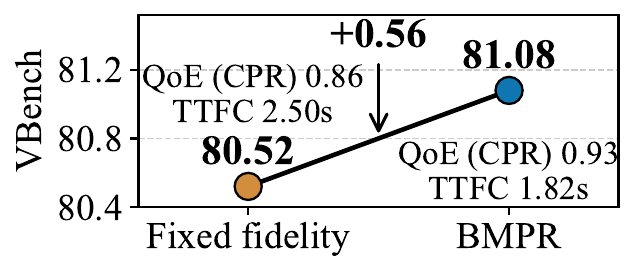}
\captionof{figure}{BMPR versus fixed-level switching.}
\vspace{-0.2in}
\label{fig:p14}
\end{minipage}
\hfill
\begin{minipage}[t]{0.46\linewidth}
\centering
\includegraphics[width=\linewidth]
{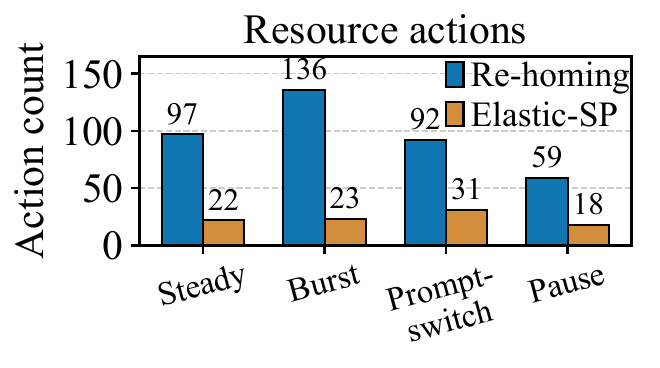}
\vspace{-0.2in}
\captionof{figure}{Trigger counts of re-homing and elastic SP.}
\label{fig:p17}
\end{minipage}

\end{figure}

\textbf{Resource-allocation decision analysis.}
Figure~\ref{fig:p17} shows that SlackServe triggers different decisions under different load patterns. Burst causes the most re-homings (136), as synchronized arrivals create cross-worker slack imbalance. Prompt-switch causes fewer re-homings (92) but more elastic-SP events (31), since condition changes create isolated low-slack chunks that require short-term expansion. Pause needs the fewest decisions (59 re-homings and 18 elastic-SP), because pauses let streams accumulate slack. These results show that service credit steers SlackServe toward re-homing for worker imbalance and elastic SP for tail recovery.

\begin{figure}[t]
\centering
\includegraphics[width=\linewidth]{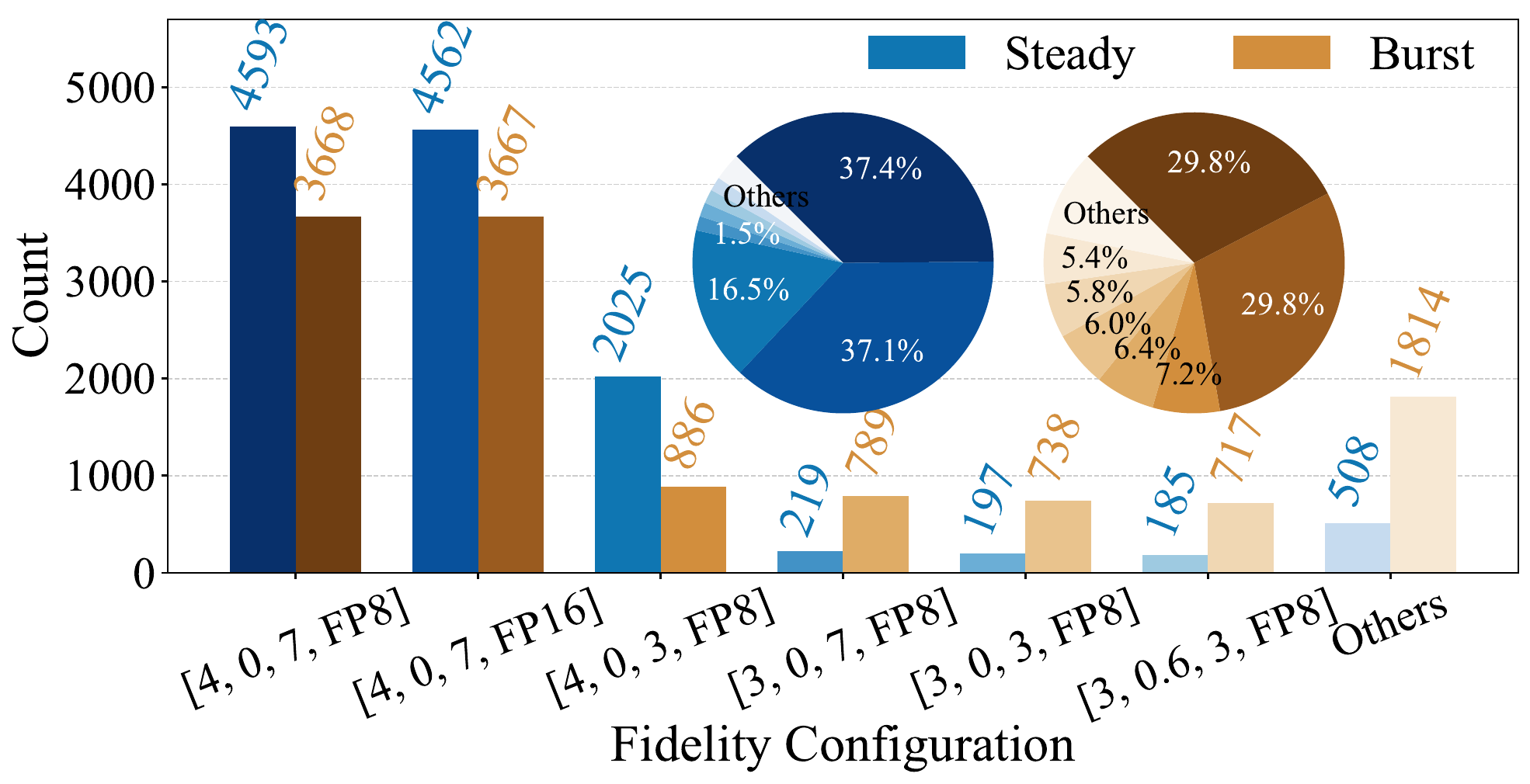}
\vspace{-0.2in}
\caption{Selected fidelity configs under Steady and Burst.}
\label{fig:p18}
\end{figure}

\textbf{Fidelity-selection analysis.}
Figure~\ref{fig:p18} compares the fidelity configurations selected by BMPR under Steady and Burst workloads. The top five configurations account for 94.1\% of selections under Steady and 79.4\% under Burst, showing that BMPR does not frequently oscillate across the full configuration space. Instead, it repeatedly selects a small set of Pareto-efficient configurations that provide speed--quality tradeoffs. The distribution also shifts with workload pressure. Under Steady load, where playout slack is stable, BMPR selects more high-quality configurations. Under Burst load, synchronized arrivals reduce slack and push more chunks toward faster configurations. This behavior shows that BMPR responds to runtime slack pressure while keeping selections within quality-preserving Pareto choices.

\subsection{Sensitivity analysis}
\label{subsec:sensitivity}

We evaluate whether \sys depends on tightly tuned parameters. All sweeps use Causal-Forcing on the Steady workload. Parameters outside the sweep use their default values. Results are shown in Table~\ref{tab:sensitivity}.

\begin{table}[t]
\centering
\caption{Sensitivity analysis under the Steady workload.}
\vspace{-0.1in}
\label{tab:sensitivity}
\small
\setlength{\tabcolsep}{6pt}
\renewcommand{\arraystretch}{1.02}
\begin{tabular}{@{}cccc@{\hspace{0.45em}}cccc@{}}
\toprule
\multicolumn{4}{c}{URGENT/RELAXED threshold} &
\multicolumn{4}{c}{Arrival rate} \\
\cmidrule(lr){1-4}\cmidrule(lr){5-8}
$\alpha$ & QoE & TTFC & VBench &
Rate & QoE & TTFC & VBench \\
\midrule
1.0 & 0.884 & 1.59 & 80.82 &
0.6 & 0.998 & 0.81 & 81.25 \\
1.5 & 0.925 & 1.77 & 81.11 &
1.0$^\dagger$ & 0.932 & 1.82 & 81.08 \\
2.0$^\dagger$ & 0.932 & 1.82 & 81.08 &
1.4 & 0.851 & 3.92 & 80.95 \\
3.0 & 0.928 & 2.10 & 80.93 &
1.8 & 0.794 & 6.92 & 80.67 \\
4.0 & 0.869 & 1.76 & 80.95 &
2.2 & 0.733 & 8.19 & 80.49 \\
\bottomrule
\end{tabular}
\vspace{0.15em}
\caption*{\small $^\dagger$Default setting. TTFC is in seconds, and Rate is in streams/s.}
\vspace{-0.35in}
\end{table}

\textbf{URGENT/RELAXED threshold.}
\sys classifies a stream as URGENT when $C_u < \alpha T_u$ and RELAXED when $C_u > 2\alpha T_u$. We sweep $\alpha \in \{1, 1.5, 2, 3, 4\}$, and find that QoE is stable for $\alpha \in \{1.5, 3.0\}$: QoE varies only from 0.925 to 0.932 and 0.928, while TTFC remains within 1.77--2.10s. This plateau shows that \sys is insensitive to the exact urgency threshold. At the extremes, $\alpha=1$ marks streams as URGENT too late, causing more elastic-SP recovery and tail stalls, whereas $\alpha=4$ over-classifies streams as urgent, increasing unnecessary migration. QoE drops to 0.869, with only a 0.16\% quality decrease. We therefore use $\alpha=2$, which lies in the middle of the stable region.

\textbf{Arrival rate.}
We sweep the Steady arrival rate from 0.6 to 2.2 streams/s. As load increases, QoE decreases from 0.998 to 0.733, and TTFC rises from 0.81s to 8.19s. The degradation is gradual rather than abrupt, showing that the controller sheds pressure progressively through priority scheduling, re-homing, elastic SP, and BMPR instead of failing at a narrow saturation point. Quality decreases moderately by 0.94\% at the heaviest load, indicating that BMPR bounds quality loss even when more chunks enter low-slack states. Even at 2.2 streams/s, \sys achieves 0.733 QoE, higher than the Steady QoE of all baselines in Figure~\ref{fig:p12}. This suggests that the gains come from the closed-loop slack-driven design rather than from tuning to the default admission rate.

\subsection{Other experiments}
We further quantify the scalability of the Control Plane and migration overhead in Appendix~\ref{app:overhead}. Overall, these experiments confirm that \sys’s \textit{Control Plane} incurs negligible runtime cost, and that asynchronous state migration keeps communication overhead largely off the critical path.

\section{Related Work}
\label{sec:relatedwork}
\textbf{LLM and DiT serving.}
LLM serving systems improve throughput and latency through scheduling, KV-cache management, batching, and parallelism~\cite{orca, vllm, sarathi-serve, distserve, loongserve}. Text-streaming serving systems, such as Andes~\cite{liu2024andes} and TokenFlow~\cite{chen2026tokenflow}, further improve perceived user experience through preemptive scheduling and token-level streaming. However, their objectives target token delivery, such as reducing token wait time or improving token-level smoothness, rather than generated-video playout continuity. Unlike text tokens, generated video chunks have explicit playout deadlines and objectively observable stall events. Therefore, these mechanisms do not directly address real-time video generation serving. Diffusion serving systems similarly optimize denoising workloads through scheduling and resource management~\cite{katz, tridentserve,diffserve,tetriserve}. Among them, TridentServe~\cite{tridentserve} dynamically adjusts parallelism to satisfy diffusion-service SLOs, while StreamDiffusionV2~\cite{streamdiffusionv2} targets interactive video generation through rolling KV cache, batching, and pipeline orchestration. However, these systems optimize request-level latency, throughput, or FPS rather than the continuously evolving playout slack of each video stream. \sys instead focuses on preserving playout continuity under streaming generation.

\textbf{Dynamic fidelity configuration and efficient video generation.}
A complementary line of work exposes speed--quality tradeoffs for video generation through model-level and kernel-level acceleration. Common techniques include directly reducing denoising steps, with distillation or consistency models further improving quality at low step counts~\cite{dmd,wang2023videolcm}, sparse attention~\cite{lightforcing,svg,adaspa}, KV-window reduction and autoregressive generation~\cite{rollingsink, rollingforcing}, and low-precision quantization~\cite{sageattention, sageattention2}. Other approaches, such as caching~\cite{flexcache,adacache}, pruning~\cite{tokenmerging}, and model cascades~\cite{diffserve}, further reduce generation cost. These methods provide different speed--quality tradeoffs. Our work focuses on the four common fidelity knobs above because they can be adjusted online at chunk boundaries and expose practical latency--quality tradeoffs for real-time AR-DiT serving.

\section{Conclusion and Future Work}

AR-DiT-based real-time video generation shifts the serving objective from completing a request quickly to keeping each stream ahead of its playout timeline. This paper presents \sys, a playout-slack-driven serving system for AR-DiT-based real-time video generation. \sys uses service credit to guide both cross-stream resource reallocation and within-stream fidelity selection. Across streams, it redirects resources to urgent streams through three-tier priority queues, re-homing, and elastic SP. Within each stream, BMPR selects Pareto-optimal fidelity configurations under a playout-slack budget and a quality floor. 

As future work, we plan to extend \sys to multi-task and multi-resolution serving, where streams may have different generation objectives, resolutions, and playout rates. While \sys is, to our knowledge, the first step toward playout-slack-driven serving for streaming video generation, it still relies on profiled frontiers and heuristic control policies, which we plan to refine.

\bibliographystyle{plain}
\bibliography{sample-base}

\newpage
\onecolumn
\appendix

\section{Fidelity-Configuration Space and Pareto Frontier in BMPR}
\label{app:profile}

This appendix details the candidate fidelity-configuration space and the Pareto frontier used by Bi-Modal Pareto Routing (\S\ref{sec:precision}).

\textbf{Fidelity-configuration space.} \sys exposes the four fidelity knobs of \S\ref{sec:backvideogen}: denoising steps $S \in \{2,3,4\}$, attention sparsity $\rho \in \{0,0.6,0.7,0.8,0.9\}$, KV-window size $W \in \{1,3,7\}$ chunks, and quantization mode $Q \in \{\mathrm{FP16},\mathrm{FP8}\}$. Their Cartesian product yields $3\times5\times3\times2 = 90$ candidate fidelity configurations $\mathrm{cfg}=(S,\rho,W,Q)$, with $(4,0,7,\mathrm{FP16})$ as the highest-quality reference.

\begin{figure*}[t!]
\centering

\begin{minipage}[t]{0.49\textwidth}
\centering
\includegraphics[width=\linewidth]{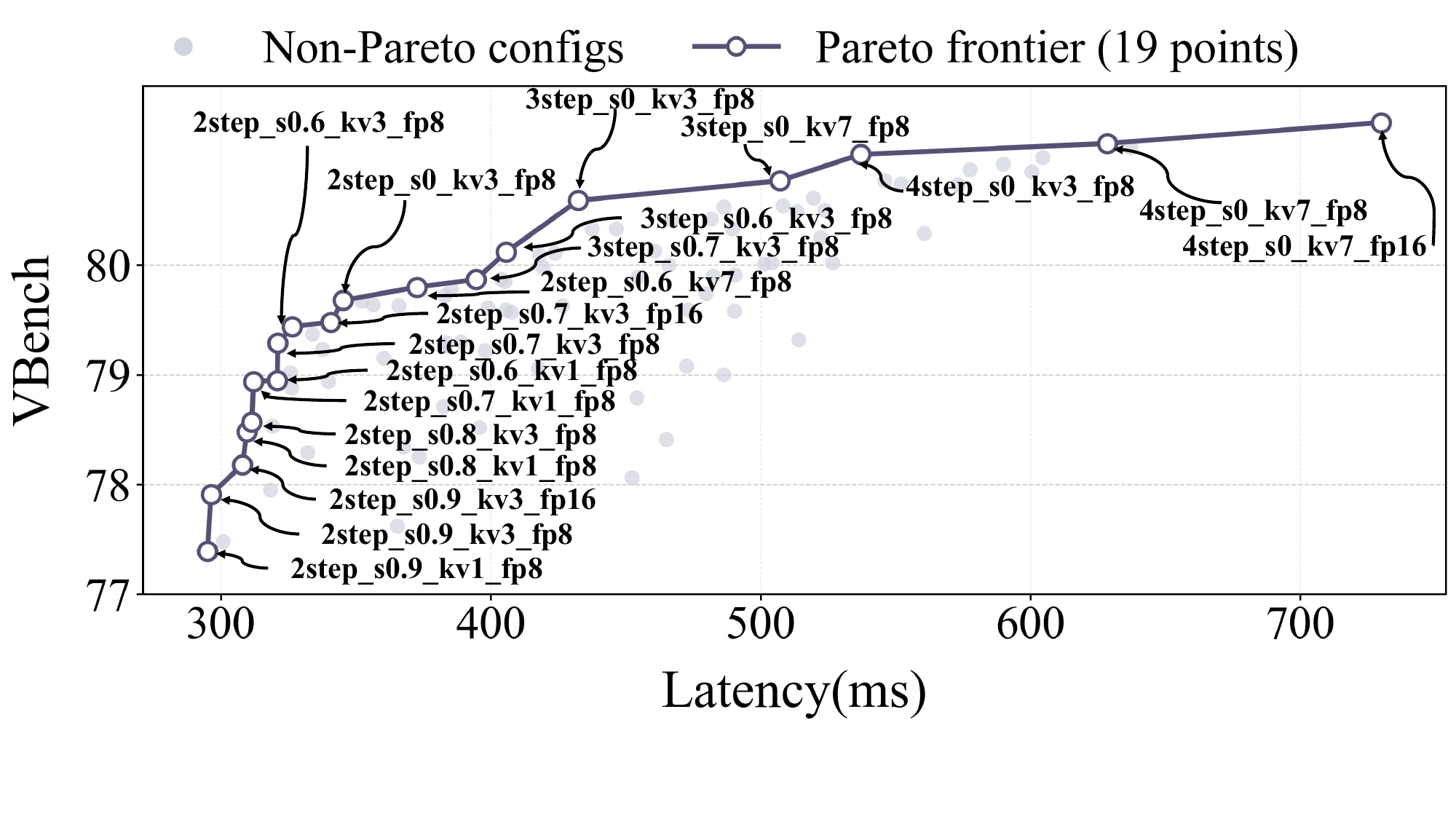}
\caption{Pareto frontier for Causal-Forcing.}
\label{fig:pareto-causal}
\end{minipage}
\hfill
\begin{minipage}[t]{0.49\textwidth}
\centering
\includegraphics[width=\linewidth]{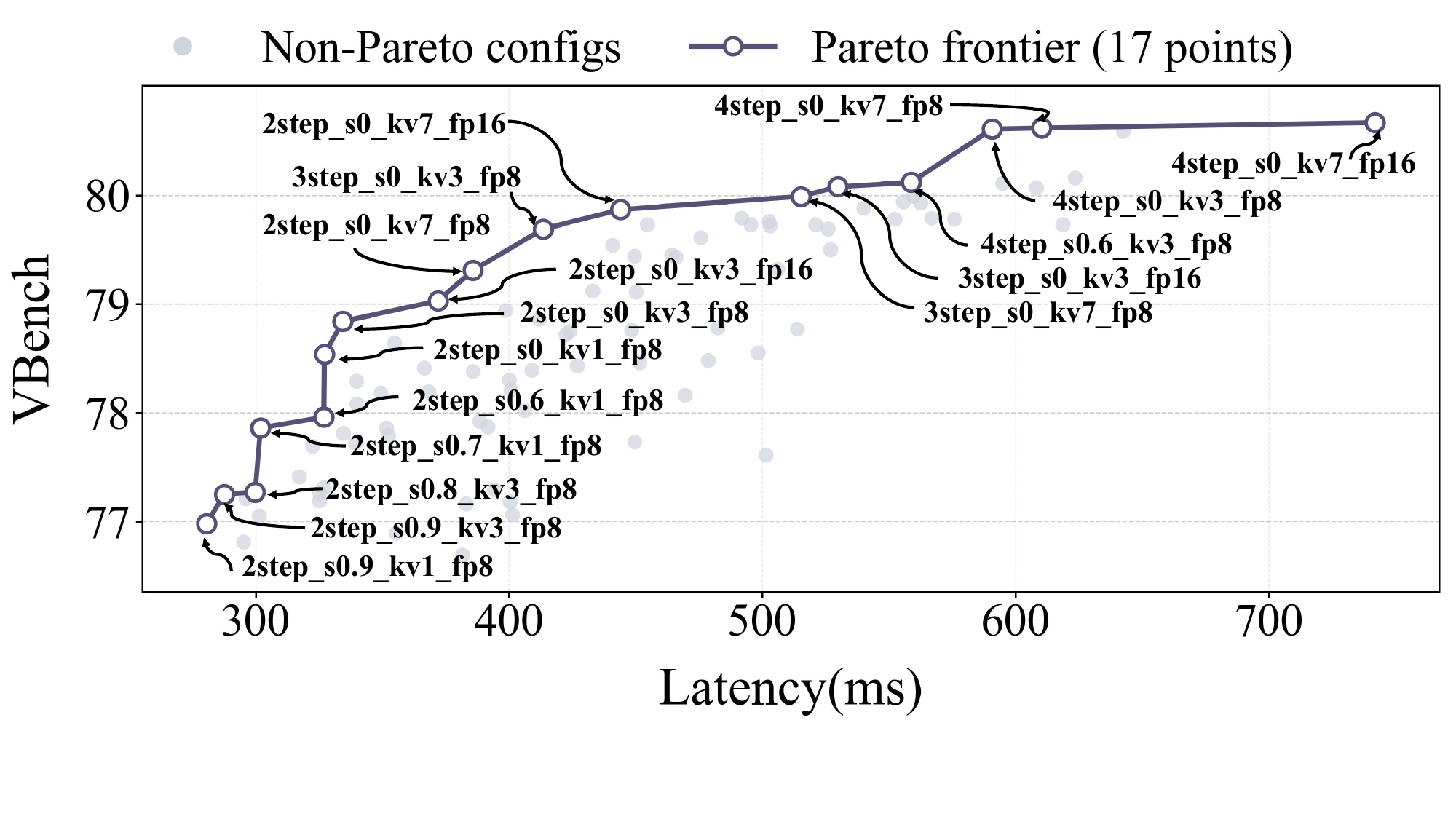}
\caption{Pareto frontier for Self-Forcing.}
\label{fig:pareto-self}
\end{minipage}

\end{figure*}

\textbf{Pareto frontier and quality floor.}
We profile every candidate's fidelity configuration offline over the VBench prompt set. For each fidelity configuration, we measure its average per-chunk generation latency $L$ in ms/chunk and obtain its quality score using VBench.
From the profiled pairs, \sys discards every fidelity configuration dominated in the $(\mathrm{latency}, \mathrm{quality})$ plane (\S\ref{sec:bmpr}) and keeps the non-dominated set as the empirical Pareto frontier $F$. Figure~\ref{fig:pareto-causal} and Figure~\ref{fig:pareto-self} show the Pareto frontiers of Causal-Forcing and Self-Forcing, together with the median quality value $Q_{\mathrm{floor}}$.
BMPR routes only within $F$ and never selects a fidelity configuration whose quality is below $Q_{\mathrm{floor}}$.

\section{Workload Details}
\label{app:workload}

This appendix details the five workloads used in \S\ref{subsec:mainresult}. Unless otherwise noted, all workloads share the same per-stream settings, with prompts drawn from VBench, which contains 946 samples. Each stream's target length is sampled uniformly from $\{81, 129, 161, 241\}$ frames, corresponding to about 5--15\,s at 16\,fps. Video is generated at 480p with three latent frames per chunk, and the playout rate is 16\,fps. Here stream lengths are counted in pixel-space frames, whereas a chunk is defined in latent frames: the two are related by the VAE temporal compression factor, which is about $4\times$ for the AR-DiTs we use, so a chunk of three latent frames corresponds to roughly twelve pixel frames (about $0.75$\,s of playout at 16\,fps).

\textbf{Steady.} Steady is the default workload. Streams arrive according to a Poisson process with rate $\lambda = 1$~stream/s, yielding 946 streams per run. Steady measures system behavior under a stationary load and serves as the base arrival process for the other synthetic workloads.

\textbf{Burst.} Burst extends Steady by introducing synchronized arrivals while preserving the total number of streams. We fix the number of burst points to three and place them at the 20\%, 50\%, and 80\% progress points of the Steady arrival sequence. At each burst point, 10\% of all streams are reassigned to arrive simultaneously at that timestamp. This construction keeps the total number of streams unchanged while creating short-lived overloads that model flash crowds, hot events, or batches of users joining at once. The resulting transient load spikes stress cross-worker re-homing (\S\ref{sec:rehome}) and test whether the system can quickly redistribute work after synchronized arrivals.

\textbf{Prompt-switch.} Prompt-switch extends Steady by injecting prompt-switch signals into each stream. Streams with target lengths of 81 frames receive one switch, streams with target lengths of 129 or 161 frames receive two switches, and streams with target lengths of 241 frames receive three switches. The switch positions are sampled uniformly at random within each stream. A prompt switch represents a user changing the generation condition and then continuing generation under the new condition; chunks buffered under the old condition are no longer useful, so the stream's playout slack is reset to the initial TTFC. To preserve the semantic continuity required by VBench evaluation, we do not actually change the text prompt. Instead, we inject a playout-slack reset signal that captures the scheduling effect of a prompt switch without altering the generated content. This workload creates isolated low-slack regions within otherwise normal streams, stressing per-chunk fidelity selection (\S\ref{sec:precision}) and elastic SP (\S\ref{sec:elastic}).

\textbf{Pause.} Pause extends Steady by injecting client-side pause signals into each stream. Streams with target lengths of 81 frames receive one pause, streams with target lengths of 129 or 161 frames receive two pauses, and streams with target lengths of 241 frames receive three pauses. Pause positions are sampled uniformly at random within each stream. Each pause lasts for 20\% of the stream's target duration. During a pause, client-side playout halts while generation may continue, allowing playout slack to accumulate. This workload models user behaviors such as temporary viewing pauses or interactions that delay consumption. Compared with the other synthetic workloads, Pause is less adversarial in terms of instantaneous load, but it tests whether \sys can preserve and exploit accumulated slack rather than spending it uniformly.

\textbf{Proprietary trace.} The proprietary trace is a realistic enterprise trace with mixed arrival patterns, including interleaved steady periods, bursts, and idle gaps. We use its arrival pattern as the external workload shape and uniformly subsample the trace arrival points to match our cluster scale. To make the trace compatible with our video-generation benchmark, we replace the original prompts and target lengths with VBench prompts and sample each stream's target length uniformly from $\{81,129,161,241\}$ frames. This workload captures more complex real-world dynamics than the synthetic traces, combining non-stationary arrivals with heterogeneous stream lengths.

\section{Additional Implementation Details}
\label{app:impl-details}

This appendix provides additional implementation details of SlackServe's
control loop, re-homing policy, elastic sequence parallelism, and
sequence-parallel KV-state transfer. These details complement the design
described in \S\ref{sec:overview} and \S\ref{sec:resource}.

\subsection{High-Level Control Workflow}
\label{app:control-workflow}

Algorithm~\ref{alg:workflow} summarizes the high-level workflow of
SlackServe. The system is event-driven. New stream arrivals are handled by
admission control, periodic control ticks update fidelity and resource-allocation
decisions, and step/chunk boundaries serve as safe points where published
decisions take effect.

At each control tick, SlackServe first selects the next-chunk fidelity
configuration for each active stream using BMPR. It then recomputes service
credit under the selected configuration and updates the stream's urgency tier.
Based on the updated tiers, the controller generates a bounded re-homing plan
and an elastic-SP plan, composes these decisions, and publishes them to the
execution and state planes. State transfers are issued asynchronously and do not
block the control loop. At safe boundaries, a worker applies the latest
published decision and dispatches the next stream from its priority queue. Fidelity change and re-homing are applied only at chunk boundaries, while priority preemption and elastic SP decisions may take effect at step or chunk boundaries.

\begin{algorithm}[t]
\caption{High-Level Workflow of \sys}
\label{alg:workflow}
\begin{algorithmic}[1]
\Require stream requests $R$, cluster of workers $C$
\State State: per-stream slack, service credit, fidelity configuration, and per-worker priority queues
\While{system runs}
  \If{a new stream request $s$ arrives}
    \State $home \gets \text{ChooseHomeWorker}(s, C)$
    \State $s.slack_{\mathrm{init}} \gets 4 \times \text{EstimatedFirstChunkTime}(s)$
    \State $\text{Admit}(s, home)$
  \ElsIf{a control tick fires}
    \For{each active stream $u$}
      \State $u.cfg_{\mathrm{next}} \gets \text{BMPRSelect}(u)$
      \State $u.credit \gets \text{ComputeCredit}(u, u.cfg_{\mathrm{next}})$
      \State $u.tier \gets \text{ClassifyTier}(u.credit, u.cfg_{\mathrm{next}})$
    \EndFor
    \State $\text{UpdatePriorityQueues}(streams, C)$
    \State $P_{\mathrm{re}} \gets \text{Rehoming}(streams, C)$
    \State $P_{\mathrm{sp}} \gets \text{ElasticSP}(streams, C)$
    \State $\text{Async}(\text{StatePlane.Transfer}(P_{\mathrm{re}}, P_{\mathrm{sp}}))$
  \ElsIf{a step or chunk boundary occurs on worker $w$}
    \State $\text{ApplyDecision}(w)$
    \If{the boundary is a chunk boundary}
      \State $\text{ApplyNextFidelityConfig}(w)$
    \EndIf
    \State $\text{DispatchNext}(w.priority\_queue)$
  \EndIf
\EndWhile
\end{algorithmic}
\end{algorithm}

\subsection{Details of Re-homing}
\label{app:rehoming-details}

Re-homing is intentionally conservative. Although moving an urgent stream to a
slack-rich worker can reduce stalls, frequent migration can consume bandwidth,
increase KV-transfer pressure, and cause oscillation. SlackServe therefore uses
three safeguards.

First, each migrated stream enters a cooldown period of 60 seconds. A stream in
cooldown is not eligible for another re-homing decision, even if it later becomes
URGENT again. This prevents repeated back-and-forth movement of the same stream
under transient slack fluctuations.

Second, SlackServe bounds the number of migrations generated at each control
tick. Each sender worker can migrate out at most two streams per tick, and each
receiver worker can accept at most one migrated stream per tick. These caps
limit transfer bursts and keep asynchronous KV migration from interfering with
foreground chunk generation.

Third, SlackServe prefers intra-node migration before cross-node migration.
Cross-node re-homing is used only when no intra-node receiver is available and
the sender remains URGENT-heavy. In our default configuration, the re-homing
planner selects receivers with no URGENT or NORMAL streams, so that migrated
streams are placed only on workers with sufficient slack headroom. These
constraints make re-homing stable and reduce migration-induced oscillation.

\subsection{Elastic Sequence Parallelism Policy}
\label{app:elastic-sp-details}
SlackServe pre-initializes all candidate intra-node SP2 groups before serving, so elastic SP does not create communication groups at runtime.
Elastic SP is used only as a last-resort recovery mechanism for streams whose
service credit is negative. In the default configuration, SlackServe restricts
elastic SP to an SP degree of at most two. That is, an urgent stream can borrow
at most one donor worker. The donor must be a RELAXED worker in the same node as
the urgent stream's current worker. If no such donor exists, elastic SP is not
triggered for that stream.

This conservative policy is motivated by two practical considerations. First,
larger SP degrees provide diminishing returns for the chunk sizes used in our
streaming setting, because the additional collective-communication overhead can
offset the reduced per-GPU computation. Second, cross-node elastic SP introduces
substantially higher communication cost and more complex state synchronization.
Restricting elastic SP to intra-node SP2 keeps the mechanism predictable and
prevents recovery actions from degrading cluster-wide throughput.

When a donor is selected, SlackServe switches the urgent stream to the
corresponding pre-initialized intra-node SP2 group. The donor is released at the
next safe boundary once the stream's service credit recovers to the NORMAL tier,
after which the stream switches back to its default SP1 execution group. Donor selection is
credit-aware: SlackServe chooses the highest-credit RELAXED worker in the same
node, so that borrowing compute is least likely to push the donor's local streams
toward future stalls.

\subsection{Sequence-Parallel KV-State Transfer}
\label{app:sp-kv-transfer}

SlackServe implements sequence parallelism following the Ulysses-style
partitioning strategy, where attention heads are partitioned across workers in
the SP group. This partitioning also determines how KV state is redistributed
during elastic SP. When a stream switches from SP1 to a pre-initialized SP2 group, SlackServe
migrates only the KV pages corresponding to the head partition assigned to the
donor worker.

This head-partition-aware transfer avoids an additional all-to-all
redistribution of the KV cache. The State Plane directly transfers the required
paged KV ranges to the destination worker according to the target SP layout.

\section{Control and State-Plane Overheads}
\label{app:overhead}

This appendix evaluates the overhead introduced by SlackServe's
control and state planes. We focus on three questions: (1) whether the
global controller can make decisions fast enough as the number of active
streams increases, (2) how long KV-state transfers take when resource
reallocation triggers state movement, and (3) how much of the transfer
time remains on the critical path after asynchronous layer-wise streaming.

\subsection{Controller Scalability}
\label{app:controller-overhead}

At each control tick, the Control Plane updates stream slack, selects
per-chunk fidelity configurations, reorders per-worker queues, and
generates re-homing and elastic-SP decisions. Let \(U\) be the number of
active streams, \(G\) the number of workers, and \(|F|\) the size of the
profiled Pareto frontier used by BMPR. The controller performs four main
operations. First, it computes service credits for all streams in
\(O(U)\). Second, BMPR selects the next fidelity configuration for each
stream by scanning the Pareto frontier, which costs \(O(U|F|)\). In our
implementation, \(|F|\) is small and fixed after offline profiling. Third,
three-tier queue construction and within-tier ordering cost $O\!\left(\sum_{g=1}^{G} U_g \log U_g\right)$,
where \(U_g\) is the number of streams assigned to worker \(g\). Fourth,
re-homing and elastic-SP planning operate over workers and urgent streams.
With bounded per-tick migration caps, their cost is linear in the number
of candidate urgent streams plus \(O(G^2)\) for sender--receiver matching.
Overall, for a fixed worker count, the controller scales near-linearly
with the number of active streams.

We measure the average end-to-end controller latency on the same 16-GPU
testbed used in the main evaluation. To stress the controller without
changing the GPU cluster size, we replay controller states with
64--1024 active streams and keep the same 3-second control interval as in
the main system. Table~\ref{tab:controller-overhead} shows that even with
1024 active streams, one control tick takes 39.6 ms on average, which is
only 1.32\% of the default control interval. This overhead is small
compared with chunk-generation time and does not affect the critical path
of video generation.

\begin{table}[t]
\centering
\caption{\textit{Control Plane} scalability on the 16-GPU testbed. We report the
average end-to-end time of one control tick, including slack update,
fidelity selection, and resource-reallocation planning.}
\label{tab:controller-overhead}
\begin{tabular}{rcc}
\toprule
\# Active streams & Avg. control time (ms) & Fraction of 3s tick \\
\midrule
64   & 9.1  & 0.30\% \\
128  & 10.8  & 0.36\% \\
256  & 13.7 & 0.46\% \\
512  & 20.4 & 0.68\% \\
1024 & 39.6 & 1.32\% \\
\bottomrule
\end{tabular}
\end{table}

\subsection{State-Plane Transfer Overhead}
\label{app:state-transfer-overhead}

SlackServe moves KV state when re-homing or elastic SP changes the worker
set of a stream. We measure all completed KV transfers on the Steady
workload. Table~\ref{tab:state-plane-overhead} summarizes both the raw
KV-transfer latency and the residual waiting time observed by affected
dispatches.

Most transfers complete within tens of milliseconds because SlackServe
prefers intra-node re-homing and issues transfers asynchronously. The
tail comes from cross-node transfers, larger KV windows, and runtime
overheads such as page lookup, transfer submission, CUDA-event
synchronization, and layer-wise transfer scheduling. Across all transfer
events, the average transfer time is 31.8 ms and the P95 transfer time is
118.4 ms.

Residual waiting is much smaller than raw transfer latency. A migrating
stream is reinserted once its first-layer KV pages are ready, while later
layers continue transferring in the background. As a result, most affected
dispatches wait for less than 5 ms. The average residual wait is 4.4 ms,
and the P95 residual wait is 16.6 ms. Only 13.8\% of the average transfer
latency remains on the critical path.

\begin{table*}[t]
\centering
\caption{\textit{State Plane} overheads on the Steady workload. KV-transfer
latency is mostly hidden by asynchronous layer-wise streaming.}
\label{tab:state-plane-overhead}
\begin{minipage}[t]{0.48\textwidth}
\centering
\caption*{(a) KV transfer time}
\begin{tabular}{lrr}
\toprule
Time range (ms) & Events & Fraction \\
\midrule
0--5     & 8  & 6.7\%  \\
5--10    & 18 & 15.0\% \\
10--15   & 24 & 20.0\% \\
15--20   & 21 & 17.5\% \\
20--30   & 17 & 14.2\% \\
30--40   & 11 & 9.2\%  \\
40--60   & 8  & 6.7\%  \\
60--80   & 5  & 4.2\%  \\
80--120  & 5  & 4.2\%  \\
120+     & 3  & 2.5\%  \\
\midrule
Avg. & \multicolumn{2}{r}{31.8 ms} \\
P95  & \multicolumn{2}{r}{118.4 ms} \\
\bottomrule
\end{tabular}
\end{minipage}
\hfill
\begin{minipage}[t]{0.48\textwidth}
\centering
\caption*{(b) Residual dispatch wait}
\begin{tabular}{lrr}
\toprule
Time range (ms) & Events & Fraction \\
\midrule
0--5    & 82 & 68.3\% \\
5--10   & 22 & 18.3\% \\
10--15  & 9  & 7.5\%  \\
15--20  & 4  & 3.3\%  \\
20--25  & 2  & 1.7\%  \\
25+     & 1  & 0.8\%  \\
\midrule
Avg. & \multicolumn{2}{r}{4.4 ms} \\
P95  & \multicolumn{2}{r}{16.6 ms} \\
\bottomrule
\end{tabular}
\end{minipage}
\end{table*}

\end{document}